\documentclass[letterpaper]{emulateapj}
 
\usepackage{lscape}

\def\arcsec{$\,^{\prime\prime}$~}
\def\arcmin{$\,^\prime$~}

\def\erg/cm2sec{ergs~cm$^{-2}$~s$^{-1}$}  
\def\ergcm2{ergs~cm$^{-2}$}

\newcommand{\lsim }{{\lower0.8ex\hbox{$\buildrel <\over\sim$}}}
\newcommand{\gsim }{{\lower0.8ex\hbox{$\buildrel >\over\sim$}}}

\def\aap{ A\&A}

\def\apjs{ApJ Supp}

\def\Chandra{${\it Chandra}$\ }
\def\HST{${\it HST}$}

\def\simge{\mathrel{%
   \rlap{\raise 0.511ex \hbox{$>$}}{\lower 0.511ex \hbox{$\sim$}}}}
\def\simle{\mathrel{
   \rlap{\raise 0.511ex \hbox{$<$}}{\lower 0.511ex \hbox{$\sim$}}}}

\newcommand{\Msun}{\ifmmode {M_{\odot}}\else${M_{\odot}}$\fi}
\newcommand{\Lsun}{\ifmmode {L_{\odot}}\else${L_{\odot}}$\fi}
\newcommand{\Rsun}{\ifmmode {R_{\odot}}\else${R_{\odot}}$\fi}

\shorttitle{X-ray Sources in 47 Tuc}
\shortauthors{Heinke et al.}

\begin{document}
\title{A Deep Chandra Survey of the Globular Cluster 47 Tucanae: Catalog
of Point Sources}   

\author{C. O. Heinke\altaffilmark{1,2,3}, J. E. Grindlay\altaffilmark{1},
P. D. Edmonds\altaffilmark{1}, H. N. Cohn\altaffilmark{4},
P. M. Lugger\altaffilmark{4}, F. Camilo\altaffilmark{5},
 S. Bogdanov\altaffilmark{1}, P. C. Freire\altaffilmark{6}
 } 

\altaffiltext{1}{Harvard-Smithsonian Center for Astrophysics,
60 Garden Street, Cambridge, MA  02138; cheinke@cfa.harvard.edu,
josh@cfa.harvard.edu, pedmonds@cfa.harvard.edu, sbogdanov@cfa.harvard.edu}

\altaffiltext{2}{Northwestern University, Dept. of Physics \&
  Astronomy, 2145 Sheridan Rd., Evanston, IL 60208}

\altaffiltext{3}{Lindheimer Postdoctoral Fellow}

\altaffiltext{4}{Department of Astronomy, Indiana University, Swain West 319,
Bloomington, IN 47405; cohn@indiana.edu, lugger@indiana.edu}

\altaffiltext{5}{Columbia Astrophysics Laboratory, Columbia
University, 550 West 120th Street, New York, NY 10027;
fernando@astro.columbia.edu} 

\altaffiltext{6}{NAIC, Arecibo Observatory, HC03 Box 53995, PR 00612,
USA; pfreire@naic.edu}

\slugcomment{To appear in the Astrophysical Journal}

\begin{abstract}

We have detected 300 X-ray sources within the half-mass radius
(2\farcm79) of the 
globular cluster 47 Tucanae in a deep (281 ks) \Chandra exposure.  We perform
photometry and simple spectral fitting for our detected 
sources, and construct luminosity functions, X-ray color-magnitude and
color-color diagrams.  Eighty-seven X-ray sources show variability on 
timescales from hours to years.  Thirty-one of the new X-ray sources
are identified with  
chromospherically active binaries from the catalogs
of Albrow et al.
The radial distributions of detected sources imply roughly 70 
are background sources of some kind.  The radial distribution of
the known millisecond pulsar systems is consistent with their
expected locations, due to mass segregation, if 
the average neutron star mass is $1.39\pm0.19$ \Msun.

Most source spectra are well-fit by thermal plasma models, except for
quiescent 
low-mass X-ray binaries (qLMXBs, containing accreting neutron stars)
and millisecond pulsars (MSPs).
 We identify three new candidate qLMXBs with relatively low X-ray 
luminosities.  One of the brightest cataclysmic variables (CVs, X10)
shows evidence (a 4.7 hour 
period pulsation and strong soft X-ray emission) for a 
magnetically dominated accretion flow as in AM Her systems.  Most of
the bright CVs require 
intrinsic $N_H$ columns of order $10^{21}$ cm$^{-2}$, suggesting a
possible DQ Her nature.  A group of X-ray sources associated with
bright (sub)giant stars also requires intrinsic absorption. 

By comparing the X-ray colors, luminosities, variability, and quality
of spectral fits of the detected MSPs to  
those of unidentified sources, we estimate that a total of
$\sim$25-30 MSPs exist in 47 Tuc ($<60$ at 95\% confidence), regardless
of their radio beaming 
fraction.  We estimate that the total number of neutron stars in 47 Tuc
is of order 300, reducing the discrepancy between theoretical neutron star
retention rates and observed neutron star populations in globular clusters.
Comprehensive tables of source properties and simple
spectral fits are provided electronically.

\end{abstract}


\maketitle
%----------------------------------------------------------------------

\section{Introduction}\label{s:intro}

Globular cluster cores have significantly enhanced  
dynamical interactions that produce a variety of close binary systems,
such as blue straggler stars and millisecond radio 
pulsars \citep{Clark75, Hut92, Bailyn95, Johnston92}.  
X-ray studies, especially with the \Chandra X-ray Observatory, have
been extremely effective in discovering compact binaries in globular
clusters \citep[for a review, see][]{Verbunt04}.  The brightest X-ray
sources (0.5-2.5 keV X-ray luminosity 
$L_X=10^{35.5-37}$ ergs s$^{-1}$) seen in a few (12) globular clusters are  
persistent or transient low-mass X-ray binaries (LMXBs) containing
neutron stars \citep{Grindlay84,Lewin83,Sidoli01}.  The
lower-luminosity X-ray sources ($L_X=10^{29-34}$ ergs s$^{-1}$)
include several kinds of systems with different X-ray emission
mechanisms.  These include accreting cataclysmic variables
\citep[CVs;][]{Hertz83, Cool95}, quiescent low-mass X-ray binaries
(qLMXBs) where accretion onto a neutron star is stopped or greatly reduced
\citep{Hertz83,Verbunt84,Campana98a}, millisecond radio pulsars
(MSPs), thought to be descendants of LMXBs
\citep{Saito97,Grindlay02}, and X-ray active
binaries (ABs) consisting of normal stars in tidally locked
short-period binaries, where the fast rotation rate drives increased
coronal activity \citep{Bailyn90b,Dempsey93a,Grindlay01a}.  

47 Tuc is a massive \citep[$M\sim10^6$ \Msun,][]{Pryor93} globular
cluster of relatively high stellar concentration, though not
core-collapsed \citep{Harris96}.  In this paper, we use a distance of 4.85
kpc \citep{Gratton03}, a neutral gas column of $N_H=1.3\times10^{20}$
cm$^{-2}$ \citep[from E($B$-$V$) of 0.024 and
  $N_H$/E($B$-$V$)=$5.5\times10^{21}$
cm$^{-2}$,][]{Gratton03,Predehl95,Cardelli89}, and a core radius of 
24\arcsec\ \citep{Howell00}.  It is one of the clusters with the 
highest predicted close encounter frequencies, implying a substantial
population 
of binaries whose properties have been altered by close encounters
with other stars or binaries \citep{Verbunt87,Pooley03}.  
47 Tuc has a bimodal distribution of blue stragger stars
\citep{Bailyn95}, with a high density in the core of presumably
collisional blue stragglers \citep{Ferraro01}, and another population
of probable primordial blue stragglers in the halo \citep{Ferraro04}. 
Various attempts to derive the fraction of stars in binaries in 47 Tuc
have given results from  10-18\% \citep[][hereafter AGB01]{Albrow01}
to $\lesssim$5\% \citep{Anderson97,Ivanova03}.  
Measuring the various binary populations of 47 Tuc is critical to
understanding and modeling the dynamical encounters between binaries
that lead to the production of X-ray sources in globular clusters.

47 Tuc is well-known for its large number of MSPs, which are believed
to have been produced by the evolution of LMXBs 
that were themselves produced through interactions of binaries and
neutron stars \citep{Bhattacharya91, Hut92, Rasio00}.  Twenty-two MSPs
have been discovered in 47 Tuc to date \citep[][X and Y still
unpublished]{Manchester91,Camilo00}, of which seventeen have known radio 
timing positions \citep[][ 47 Tuc-R's timing solution is not yet
 published, Freire et al.\ 2005 in preparation]{Freire01a,Freire03}, and an
additional MSP (47 Tuc-W) has a position 
determined by matching a known MSP orbital period to the period of a
variable optical star and X-ray source \citep{Edmonds02b}.
Thirteen are members of binary systems, and at least two (possibly up
to four) MSP companions have been detected using \HST\ 
\citep{Edmonds01,Edmonds02b,Edmonds03a}.  Although most of the MSPs in
47 Tuc have flux densities below the 
sensitivity of the Parkes radio telescope used to detect them, strong
interstellar scintillation effects occasionally bring their signals up 
to detectability.  Assuming that the luminosity distribution of 47 Tuc 
pulsars is similar to that of field systems, \citet{Camilo00}
estimated a total MSP population of $\sim$200 systems.
\citet{Grindlay02} detected blackbody-like thermal emission from the
surface of most of the MSPs in 47 Tuc, presumably hot polar caps.
They estimated a total MSP population in 47 Tuc of 35--90  
based on the X-ray properties of the X-ray-detected MSPs compared to
 other 47 Tuc X-ray sources.  \citet{Edmonds03b} provided evidence
that the total 47 Tuc MSP population was closer to 30--40, from optical 
identifications of X-ray sources.  
\citet{McConnell04} inferred that the observable radio MSP population in 
47 Tuc is less than 30, using observations that attempted to
detect the integrated radio flux from a population of unresolved pulsars.  Since MSPs are
highly compact, X-rays from surface hot spots will be bent by gravity to
allow observers to see $\sim$75\% of the neutron star surface
\citep{Pechenick83,Beloborodov02,Bogdanov04}. This virtually ensures
that MSPs whose  
radio beams do not intercept the Earth should still be detectable in X-rays, 
and with the lack of scintillation and eclipses due to ionized gas \citep[as
seen in the radio,][]{Camilo00}, allows more complete study of MSPs
by X-ray observations than radio, and places constraints on the radio
beaming fraction \citep{Grindlay02}.

Low-luminosity X-ray emission from 47 Tuc ($L_X\sim10^{34}$ ergs
s$^{-1}$) was first identified with
the {\it Einstein} X-ray Observatory, and suggested to be due to
cataclysmic variables \citep{Hertz83}.  Deep pointings by the {\it
  ROSAT} satellite 
measured a lower X-ray luminosity and resolved the emission into nine
sources, of which several showed variability \citep{Hasinger94, Verbunt98}.
The first \Chandra (ACIS-I) image of the cluster resolved 104
sources within a 2\arcmin by 2\farcm5 region including the core
\citep[][hereafter GHE01a]{Grindlay01a}.  A later study of the same
data extending to a radius of 4\arcmin studied the X-ray emission from
most of the MSPs with known positions, and measured the radial
distribution of the 
X-ray sources \citep{Grindlay02}.  X-ray spectra of the two brightest
X-ray sources were well-fit by hydrogen-atmosphere models for neutron
stars, consistent with the spectra of known qLMXBs \citep{Heinke03a}.  

Optical identification efforts began shortly after the discovery of
X-ray emission from 47 Tuc, but have proven difficult due to the
intense crowding at the center of 47 Tuc and the faintness of optical
counterparts.  \citet{Auriere89} suggested a possible CV counterpart, AKO 9,
to the {\it Einstein} X-ray source.  \citet{Paresce92},
\citet{Paresce94}, and \citet{Shara96} identified three CVs in 47 Tuc
through \HST\ imaging, which were later identified with {\it ROSAT}
X-ray sources \citep[][GHE01a]{Verbunt98}.
Definitive identifications became possible using \Chandra positions, 
and \HST\ data \citep[especially the exceptional GO-8267 8.3-day time series,
][]{Gilliland00},  
to search for variability and unusual colors \citep[GHE01a,][]{
Edmonds01, Edmonds02b, Edmonds02a, Edmonds03a, Edmonds03b}.  These
searches at X-ray positions produced secure optical identifications of
22 CVs, 29 ABs, one eclipsing qLMXB, and two 
 MSP companions (two additional possible MSP companions were also
 identified).  

\Chandra X-ray studies of numerous other globular clusters have given insight
into the nature and formation processes of the brighter ($L_X \gtrsim
10^{31}$ ergs s$^{-1}$) sources \citep{Pooley02a, Pooley03, Heinke03d}.
47 Tuc is an ideal target for study of the fainter X-ray sources,
including most MSPs and ABs.  This is due to its high rate of close
encounters (a product of its large mass and high density), 
relatively small distance, low reddening, reasonably 
large angular core size (reducing X-ray crowding), and excellent
multiwavelength coverage \citep[also including far-UV \HST\ spectra
  and imaging,][]{Knigge02}. 

In this paper, we describe the \Chandra X-ray data, and attempt to
understand the nature and characteristics of the faint X-ray sources.
We focus on the 2002 data, but include some analysis of the original
2000 \Chandra dataset.  
Additional papers will focus on particular aspects of the X-ray data,
individual sources, and analysis of archival and new simultaneous \HST\ 
data.  In \S\ref{s:obs}, we describe the observations and initial analysis,
including source detection, optical counterpart identifications, and
extraction of X-ray data.  In \S\ref{s:var} we describe searches for
short- and  
long-term variability.  In \S\ref{s:radial} we discuss the radial
distributions of our sources, 
and in \S\ref{s:phot} we utilize the X-ray ``color'' information. In
\S\ref{s:lf} we discuss the luminosity 
functions of the various source classes.  In \S\ref{s:spec} we
perform simple spectral fitting 
and briefly discuss some unusual objects we identify.  In
\S\ref{s:disc} 
we discuss the meaning of our results and constrain the populations of 
unidentified sources.

\section{Observations and Analysis}\label{s:obs}

The data used in this paper are from the 2000 and 2002 \Chandra
observations of the globular cluster 47 Tuc.  The 2000
observations, initially described in GHE01a, were performed
with the ACIS-I CCD array at the telescope 
focus, while the 2002 observations placed the back-illuminated ACIS-S
aimpoint at the focus for maximum low-energy sensitivity.  Five
consecutive observations were performed in 2000, as listed in 
Table~\ref{tab:obs}, with three short observations interleaved to
obtain spectra 
of bright sources with little pileup.  The 2002 observations (four
pairs of $\sim$65 ksec full-frame exposures followed by $\sim5$ ksec
1/4 subarray exposures; see Table~\ref{tab:obs}) were
designed to probe variability on a range of time scales from hours to
weeks. This was intended to allow distinctions between relatively
constant thermal MSP 
emission and flaring behavior from ABs.  The first two 2002 exposure
pairs were run consecutively for a total of 1.7 days, with the
remaining pairs following at intervals of 1.4 days and 7.4 days.

Our data analysis for all observations begins with the Level 1
 processed event lists provided by the \Chandra X-ray Center pipeline
 processing.  We used the CIAO data analysis tools\footnote{Version
3.02, http://cxc.harvard.edu/ciao/.} for initial data processing.  
 We reprocessed the 2000 observations using the CTI correction
algorithm implemented in CIAO 3.02 acis\_process\_events.  Both the 2000
and 2002 observations were reprocessed to remove the 0\farcs25 pixel
randomization added in standard processing.  We removed bad columns,
 bad pixels, and events not exhibiting one of the ``standard ASCA
 grades'' (0, 2, 3, 4, 6), and applied the good time intervals
 produced by the CXC pipeline.  We also applied the new time-dependent
 gain correction software (A. Vikhlinin, priv. comm.\footnote{See
 http://cxc.harvard.edu/contrib/alexey/tgain/tgain.html.}) to all event
 files.

We produced two sets of event lists for each
 observation for different
 analysis purposes; an imaging event list for source detection and
 positioning, and a spectral event list for use in extracting spectra
 and lightcurves.  For imaging, we removed times of high background more
 aggressively, and removed events identified with the CIAO procedure
  acis\_detect\_afterglow.
 Cosmic rays striking the detector release an initial charge that is
 generally rejected, but often release residual charge over
 10-30s.   acis\_detect\_afterglow identifies likely cosmic-ray
afterglows, which can otherwise be mistaken for true sources by
 detection algorithms.  However, the procedure also flags counts
 from real sources \citep{Feigelson02}\footnote{Also see
 http://asc.harvard.edu/ciao/threads/acisdetectafterglow/.}, so for
 extraction of spectra and lightcurves we use event lists with flagged
 events included.  

Inspection of the background
 lightcurve shows one period of significant flaring, 7.1 ks at the end
 of OBS\_ID 2736, increasing through OBS\_ID 3385.  
 A milder increase in 
background is also seen for 13 ks of OBS\_ID 2738 and OBS\_ID 3387.   
We include all of OBS\_IDs 2738 and 3387 in our analyses, since
 this gives us a higher signal-to-noise 
 ratio for our detected sources.  
For imaging analysis (where our goal is the detection of
 faint sources), we excise the  
 period of high background from OBS\_ID 2736, and do not use OBS\_ID
3385 (leaving an effective
 exposure time of 273964 s, total time). For spectral 
 and lightcurve extraction, we use all of OBS\_ID 2736, but omit
 OBS\_ID 3385 (leaving good time of 281074 s for sources within the
area covered by the subarrays, 275529 s for those outside).  
Since the total background within a typical (5 arcsec$^2$) source 
extraction area is less than 4 photons, more aggressive background 
flaring removal does not improve our signal-to-noise ratio for sources
 with more than 20 counts.  For the
2000 observations, the total good time is 72155 s. 

For this paper we confine ourselves to analysis of sources within the
half-mass radius of 47 Tuc, 2\farcm79 \citep{Harris96}.  We choose this
radius for easy comparison 
with analyses of other clusters \citep[e.g.][]{Pooley02a, Pooley02b,
Heinke03c, Pooley03, Bassa04}, and because it strikes a balance
between inclusion 
of almost all real cluster sources and exclusion of most foreground
and background sources at the flux levels we are most interested in.
 X-ray sources or MSPs associated with the cluster beyond the
half-mass radius are known in core-collapsed clusters like NGC 6752
\citep{D'Amico02} and M15 \citep{Phinney91}, and in
$\omega$ Cen \citep{Cool02b, Gendre03a}, which has not yet
dynamically relaxed.  However, in relaxed King-model clusters the
relatively massive binaries are expected to be more centrally concentrated
\citep{Verbunt88,Fregeau03}. 
The radial distribution studies of the 2000 \Chandra data by
\citet{Grindlay02}  and \citet{Edmonds03b} showed that all identified
X-ray source populations in 47 Tuc are 
highly concentrated in the core, as expected since both binaries
and single MSPs are significantly more massive than typical stars in
47 Tuc.  

\subsection{Source Detection}\label{s:detect}

We performed all data reduction separately for the 2000 and 2002
observations, since they were taken with different detectors. The CIAO
tool WAVDETECT was used to measure the positions of the three
brightest sources in different exposures.   We applied small shifts
 to align the later 2002 exposures with the first 2002 exposure
(+0\farcs028,+0\farcs023;-0\farcs012,+0\farcs063;-0\farcs026,+0\farcs041  
added to the RA and Dec of the three later exposures, with rms residuals 
of 0\farcs013, 0\farcs014 remaining).  We
reprojected the exposures in each group to match the alignment of the
first long exposure in each year, and merged each group of observations.

For the 2000 observations we performed WAVDETECT runs at full
resolution in the 0.5-2 keV and 0.5-6 keV energy bands.  For the 2002
observations we used bands of 0.3-2 and 0.3-6 keV.  For each WAVDETECT
run we selected scales of 1.0, 1.414, 2.0, and 2.828 pixels, for
optimal detection and separation of point sources near on-axis.  We
selected a threshold probability of $1\times10^{-5}$, designed to give
one false source per $10^5$ pixels.  Each WAVDETECT run (assuming the
trials and pixels are independent) should thus give of order 3.6 spurious
sources, given the 0\farcs492 pixel size and 2\farcm79 radius
searched.  (Our results tend to indicate that this is a slight
overestimate, see below).  Harder energy
bands failed to reveal more than 2 or 3 marginal additional sources
not detected in the other bands.  Inspection of
harder energy band images confirmed that the low column density to 47 Tuc and
non-accreting nature of the faintest sources (ABs and MSPs) make 
the faintest sources quite soft.  WAVDETECT is often inefficient at
separating two or more point sources less than 3\arcsec\ apart (as
noted by Feigelson et al. 2002).
Several faint sources within a few arcseconds of brighter sources can
be clearly picked out by eye, but are not identified by WAVDETECT.  We
add ten sources identified by eye to the 2002 results, and
twelve sources to the 2000 results, without reference to any
information about possible counterparts at other wavelengths.   We
also check our sourcelist 
for any spurious detections of cosmic rays missed by
acis\_detect\_afterglow (identified when more than
half of the counts from a faint source fall within a 30 second time
interval, with 
each event decreasing in energy from the last).  Several such spurious
detections are identified and eliminated in the 2000 data, including
W88 and W99 from 
the source list of GHE01a.\footnote{The acis\_detect\_afterglow
algorithm seems to detect all 
cosmic rays on the back-illuminated ACIS-S3 chip, unlike on the ACIS-I
chips.}  These spurious sources also fail to appear 
in the 2002 data, while all but three other sources detected in the
2000 data are also detected in the 2002 dataset. 
The recovery of nearly all the detected 2000 sources in the 2002 data
provides an upper limit to WAVDETECT's spurious source detection rate,
implying that probably no more than two or three of the 2002 sources
are spurious. 

We combine the results from the two
energy bands to make independent source lists for the 2000 and 2002
observations, given in Tables~\ref{tab:pos} and \ref{tab:pos2000}.
146 sources are detected in 
this way in the 2000 observations, while 300 sources are detected in
the 2002 observations.  143 of the sources are clearly detected
in both observations, while only three of the sources from the 2000
observations are not detected in the 2002 observations. Of these
three, two (W149 and W172) show only 2-3 counts, and may not be real
sources, while 
W68 \citep[identified with an optical AB counterpart,][]{Edmonds03a}
seems to have disappeared.  We retain the
W numbering scheme of GHE01a, as extended in \citet{Grindlay02} and 
\citet{Edmonds03a, Edmonds03b}, for 
the 2000 observations, extending it to cover first the remaining 2000
sources and then the additional 2002 sources. We also generate
IAU-registered positional source names, in the form CXOGLB J[xxxx.x-xxxxx]. 
We list the sources in order of decreasing total counts in both
tables.  Due to 
the iterative nature of source list construction over several papers,
the W numbering scheme does not follow clear patterns and has holes
(numbers that do not correspond to real sources within our chosen field).

\begin{figure}
\includegraphics[scale=1.1]{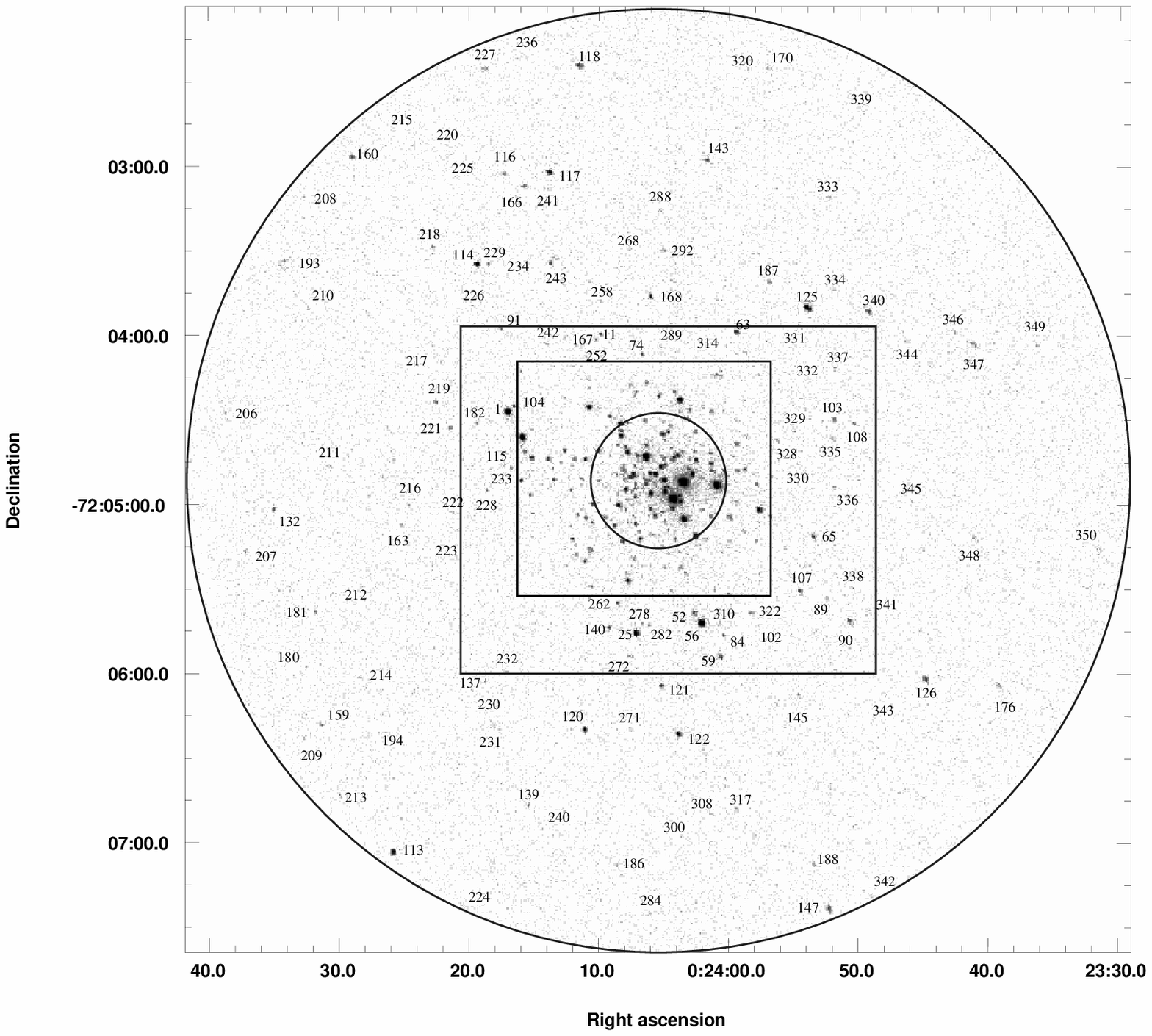}
\caption[2002_half_pub.ps]{Combined 2002 0.3-6 keV \Chandra data of
47 Tuc, out to the half-mass radius.  Circles indicate the core
(inner) and half-mass radii.  Two boxes indicate the region analyzed
in GHE01a and the (smaller) inset region shown in more detail in Figure
2.  47 Tuc X-ray sources identified in this paper outside the inset
region are labeled with their shorthand W-numbers (see Table~\ref{tab:pos}). 
} \label{fig:halfmass}
\end{figure}

\begin{figure}
\includegraphics[scale=1.05]{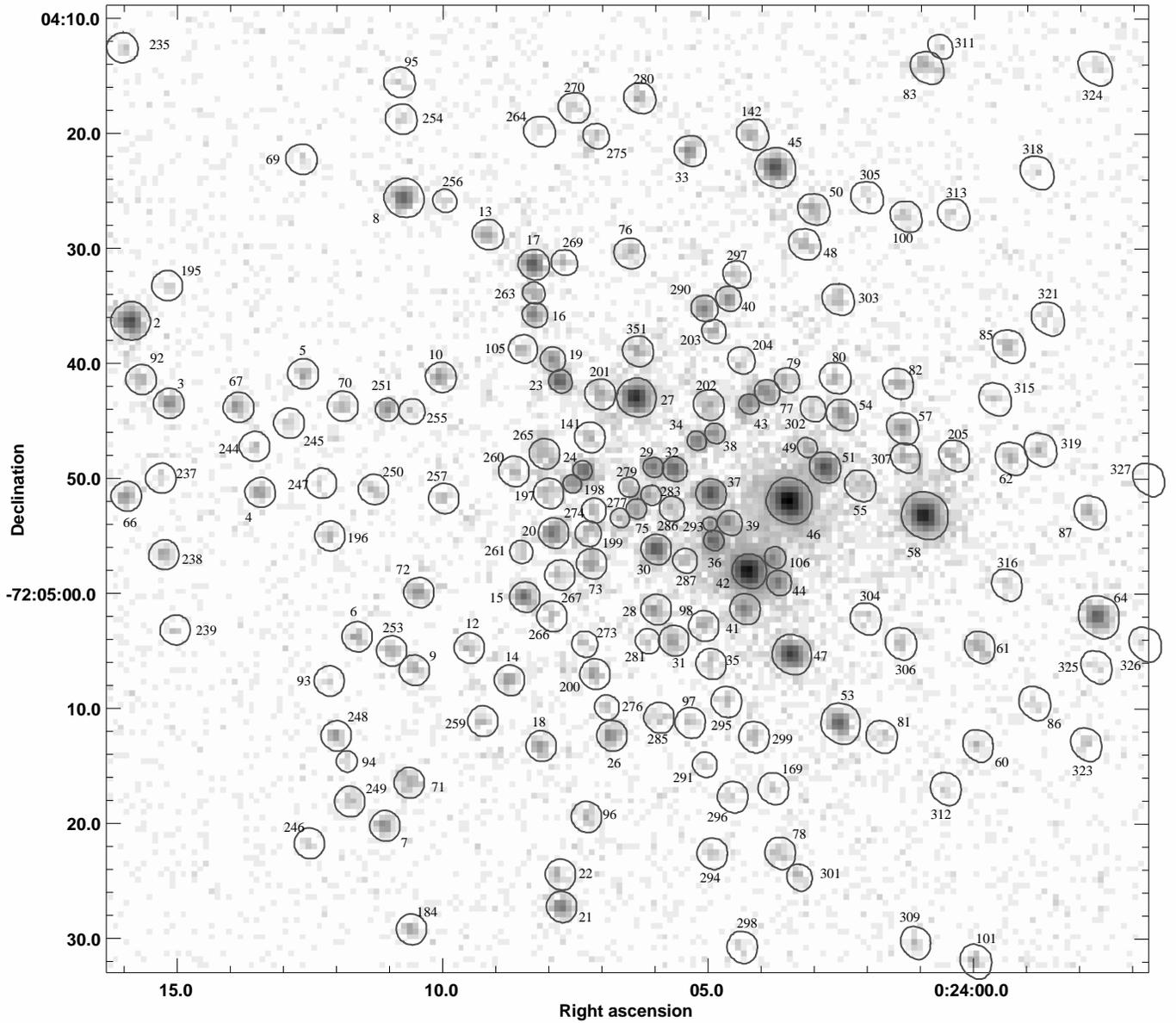}
\caption[2002_inset_pub.ps]{Combined 2002 0.3-6 keV \Chandra data of
47 Tuc, within the inset box of Figure 1. The polygons used for
extracting source events, photometry, and spectra (from the first 2002
OBS\_ID; regions for other OBS\_IDs are similar) are indicated, and labels 
indicate the W-numbers of the X-ray sources discussed in this paper.
The grayscale is logarithmic from zero to 8306 counts.
} \label{fig:inset}
\end{figure}

\begin{figure}
\includegraphics[scale=.80]{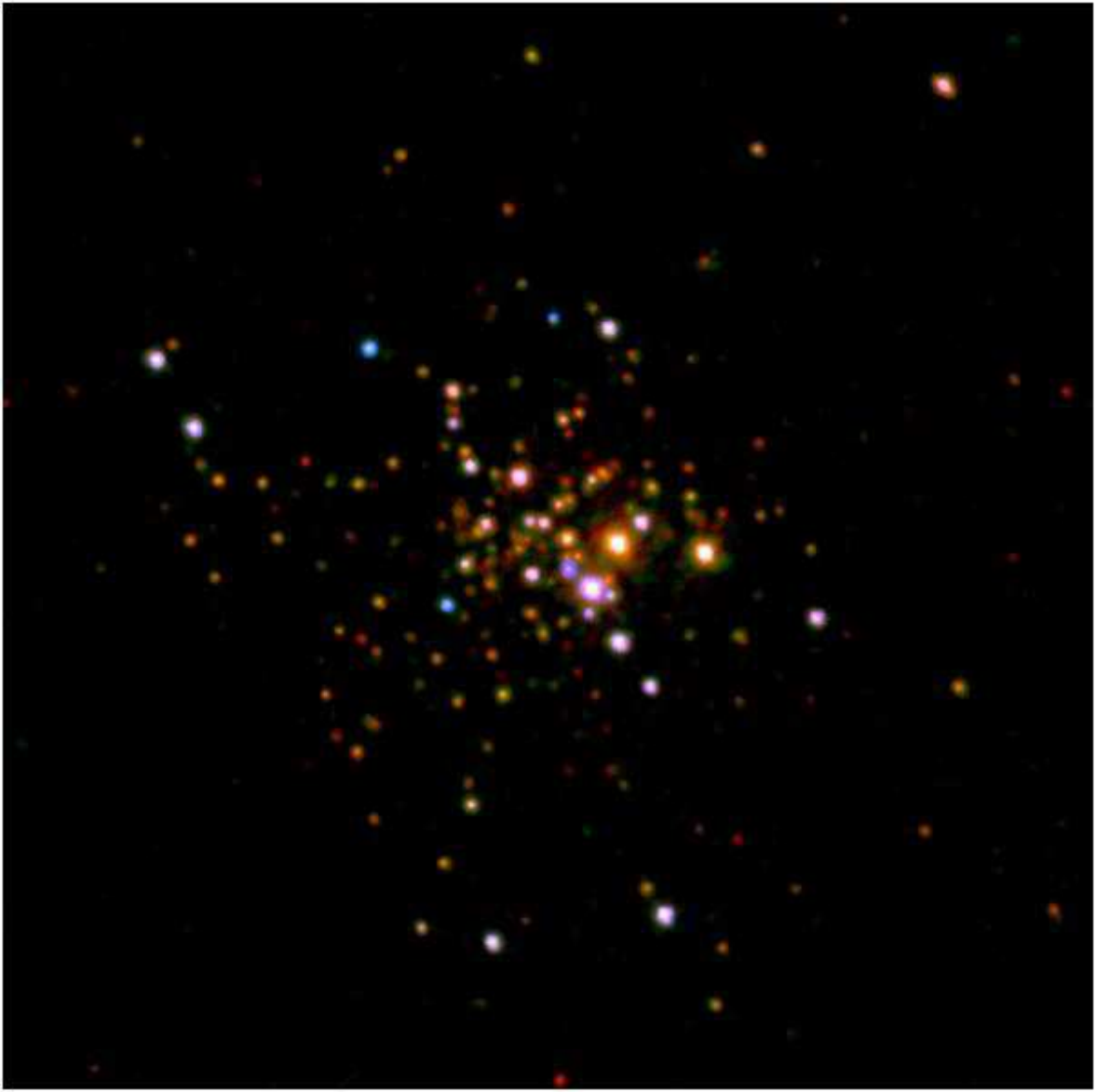}
\caption[3col-2a-ds9.ps]{Combined 2002 exposure-corrected
\Chandra data of the core region of 47 Tuc (148\arcsec square).  This
representative-color image was constructed from a 0.3-1.2 keV image
(red), a 1.2-2 keV image (green), and a 2-6 keV image (blue).  All
images were overbinned by a factor of two, smoothed using the CIAO
tool {\it csmooth}, and then combined.} \label{fig:color}
\end{figure}

In Figures~\ref{fig:halfmass} and \ref{fig:inset} we 
show the raw unbinned photon count data from the combined 2002
observations we used for source detection, in the
energy range 0.3-6 keV.  The W source
numbers are indicated in both figures.  
In Figure~\ref{fig:color} we show a
representative-color image of the 2002 data made from the 0.3-1.2, 1.2-2,
and 2-6 keV exposure-map corrected data, overbinned and smoothed on a
scale of 0\farcs74.

We align our absolute astrometry with the 17 MSPs in 47 Tuc which
have positions previously determined through radio timing studies
\citep[or \HST\  optical followup,][]{Edmonds02b} to
milliarcsecond levels \citep{Camilo00,Freire01a,Freire03}. We find
that each of the MSPs with radio timing (or optical) positions is detected, 
with only the close  pairs 47 Tuc G \& I (separation 0\farcs12) and F
\& S (separation 0\farcs74) unresolved.  \citep[We do not attempt to
separate their emission in this work; detailed analysis of the MSP
X-ray emission is performed in ][]{Bogdanov04}.  This
procedure has been described by GHE01a 
and \citet{Edmonds01,Edmonds03a} for the 2000 \Chandra observations, where
six cleanly detected MSPs were used to
boresight the \Chandra frame to the international celestial reference
frame.  We repeat this procedure for the 2002 observations.
 The rms deviations for 14 MSPs (excluding W77, 47 Tuc-F \& S) are only
0\farcs18 ($\alpha$) and 0\farcs22 ($\delta$) in the 2002 data.  After
this work was submitted for publication, a radio timing solution for
the position of 47 Tuc R became available (Freire et al.~2005, in
prep), which is offset from W198 by only $\Delta \alpha$=+0\fs03,
$\Delta \delta=-$0\farcs01 and thus a convincing identification.

\subsection{Optical Identifications}\label{s:id}

Most of the identifications for objects discussed in this work rely
upon the optical identification results of \citet{Edmonds03a,
Edmonds03b}, using the 2000 \Chandra dataset and \HST\ programs
GO-8267 (PI Gilliland) and GO-7503 (PI Meylan).  We also include an
identification by \citet{Ferraro01} 
of a bright red star as a candidate counterpart for W54, which
\citet{Edmonds03a} suggest is a possible RS CVn star. 

In addition to these identifications, we compare the variable and blue
stars  of \citet{Geffert97}, CV candidates of \citet{Knigge02}, and
binaries of AGB01 to our data.  No
additional matches are 
found with the star lists of Geffert et al.~or CV candidates of Knigge
et al., but 
the binary lists of AGB01 provide a large number of
plausible candidate counterparts, as we describe below.  

Numerous binaries in the GO-8267 \HST\ data have already been
identified as X-ray sources by \citet{Edmonds03a}.  We use the
astrometric plate solutions of \citet{Edmonds03a} to align each
\HST\ WFPC2 chip to our \Chandra frame. Then we compared the AGB01 
binaries to the positions 
of our \Chandra sources and identified those binaries which fell
within $5\sigma$ in both right ascension and declination as likely
matches (where the errors are the \Chandra WAVDETECT errors added in
quadrature with the systematic errors in the plate solutions).  In
this way we recovered every AGB01 binary previously matched to an 
 X-ray source, and identified 32 new possible matches.  These include an
alternative match for W93, the optical variable WF4-V02, that appears
closer to the \Chandra\ 
position than the marginal ID identified by \citet{Edmonds03a}. The
new matches are listed in Table~\ref{tab:ids} with their offsets from the
aligned \Chandra positions, and classifications from AGB01.  The most
discrepant new match we discuss is PC1-V21 with W287, at 1\farcs01.
This is a 2.87 $\sigma$ discrepancy, with a chance coincidence
probability of roughly 2.9\%. 

\begin{figure}
\includegraphics[scale=.80]{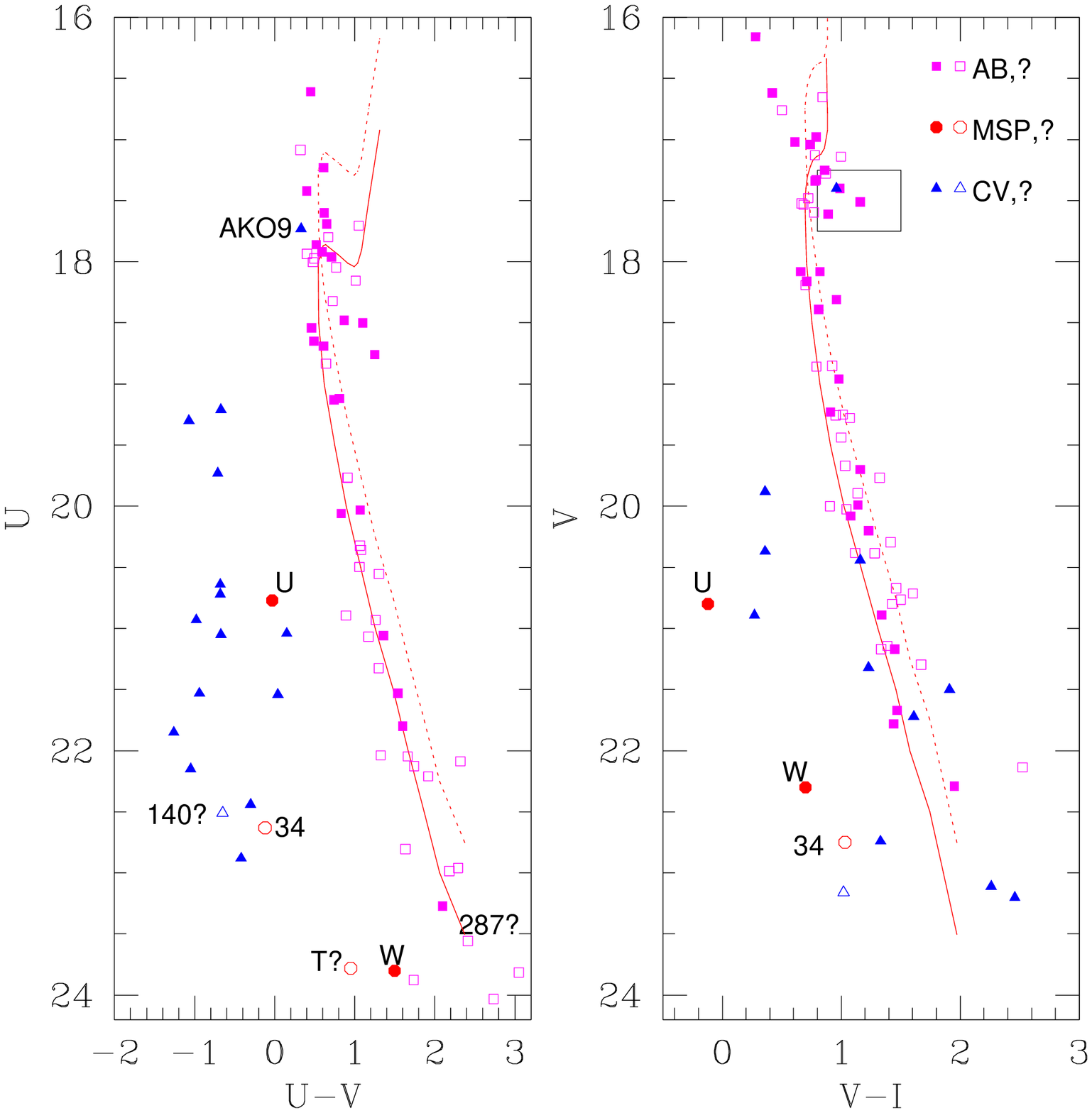}
\caption[optcmd_bw.eps]{$U$ vs.~$U-V$ and $V$ vs.~$V-I$ CMDs, showing
 only the optical counterparts for \Chandra\ sources within the GO-8267
 \HST\ field of view.  New X-ray counterpart ABs from AGB01 are
 indicated with open squares, while firm optical counterparts from
 \citet{Edmonds03a} are indicated with filled symbols.  Four
 less certain counterparts \citep[see][]{Edmonds03a} are also labeled,
 and indicated with open 
 symbols (W34--either an MSP or a CV--and the possible optical counterparts 
 to W140, 47 Tuc-T and W287).  The
 main-sequence ridge line (solid) and equal-mass binary sequence
 (dotted) are
 indicated in both CMDs, and a box in the $V$ vs.~$V-I$ CMD indicates
 the red straggler region from AGB01. (Please see the electronic
 edition of the Journal for a color version of this figure.)
}\label{fig:optcmd}
\end{figure}

  The rms total offsets for the new 
AGB01--\Chandra matches are 0\farcs42, 0\farcs21, 
0\farcs32, and 0\farcs21 for the PC, WF2, WF3, and WF4 chips
respectively (0\farcs23 for the PC chip when W287 is excluded).   
We note that the \Chandra ACIS-S array is likely to have different
small systematic astrometric errors (due to its different geometry)
compared to the ACIS-I array.  Offseting the \Chandra positions by
5\arcsec\ in right ascension and declination produced two spurious
matches in four trials, suggesting an even probability that one of our 
32 new matches is spurious.  
Figure~\ref{fig:optcmd} plots optical color-magnitude diagrams including the
new AGB01 matches, and the \citet{Edmonds03a} matches, from the 
GO-8267 \HST\ data.  It is
clear that the new AGB01 matches are similar to the previously identified
ABs, but optically fainter on average (mean $V$ magnitudes 19.3
vs. 18.7 respectively).  In the remainder of this paper we include
W287 among the candidate ABs, while the other 31 new identifications from
AGB01 are included among the confirmed ABs.  

We note that some variable stars listed by AGB01 appear to the eye to be faint
and/or confused X-ray sources that were missed by WAVDETECT.  For
completeness we list these stars here, although their selection is
subjective and may be biased.  Therefore we do not discuss them in the
rest of this 
paper.  They are PC1-V07, PC1-V20, WF2-V47, PC1-V22, PC1-V28,
PC1-V31, PC1-V33, PC1-V35, PC1-V37, PC1-V38, WF2-V14, WF2-V20,
WF3-V21, and WF4-V09.  These stars are similar in their optical
properties to the stars matched to detected X-ray sources above.  We also note
that there is no evidence for X-ray emission from the seven unusual ``blue
variables'' discussed by AGB01 and \citet{Edmonds03a,Edmonds03b};
90\% confidence upper limits for all but PC1-V36 (which lies in a confused
region) are $<1.4\times10^{29}$ ergs s$^{-1}$.

One hundred and fifty-three X-ray sources in the 2002
dataset lie within the GO-8267 \HST\ field of view. Of these X-ray
sources, 3 are
identified (by X-ray spectral fitting, see below) as qLMXBs, 12 are
identified as 13 MSPs\footnote{MSPs G and I, 0\farcs12 
  apart, are unresolved in X-rays.} by matching X-ray and radio
positions, 15 are optically identified as CVs, and 57 are optically
identified as ABs\footnote{W68, an AB 
  from the 2000 data, is not detected in the 
2002 data.}.  66 X-ray sources remain unidentified in this \HST\ field,
only 43\% of 
the total.  Further \HST\ identifications of optical counterparts,
both from the archival \HST\ datasets and from our new simultaneous
\HST\ ACS $B$, $R$, and $H\alpha$ data, are in progress.  
Details of the astrometric solution linking the new optical
identifications, all the previously identified counterparts, 
and new optical counterparts will be presented in van den Berg et
al.~(2005, in preparation). 

\subsection{Source Extraction}\label{s:extract}

For source position improvement, photometry, extraction of spectra
and lightcurves, we used the IDL (Version 5.4) tool ACIS\_EXTRACT (Broos et
al. 2002)\footnote{Version 2.7,
http://www.astro.psu.edu/xray/docs/TARA/ae\_users\_guide.html.}, 
which uses CIAO and
FTOOLS\footnote{Version 5.2,
http://heasarc.gsfc.nasa.gov/docs/software/ftools/ftools\_menu.html.}
tools, ds9 display capability\footnote{Version 2,
http://hea-www.harvard.edu/RD/ds9/doc/ref/ref.html.}, and the
TARA\footnote{http://www.astro.psu.edu/xray/docs/TARA/.} IDL
software. All spectral fitting was done in XSPEC (Arnaud 1996), version
11.2, much of it within the ACIS\_EXTRACT package using scripts
written by 
K. Getman.  Our source extraction procedure generally follows the methods
of \citet{Feigelson02} and \citet{Muno03}.

The ACIS\_EXTRACT source extraction process begins by calculating
region files for each observation designed to match a user-specified
contour level of the 
ACIS point-spread function (psf), calculated at that position with the
CIAO tool
mkpsf.  For most of our sources, we specified a 90\% encircled energy contour
(evaluated at 1.5 keV, since most of our sources are relatively soft), due to
severe crowding in the central regions. We increased this contour to
95\% for the brightest sources and for relatively bright sources
several arcminutes from the cluster center, and chose smaller regions
that did not overlap for sources very close to each other, reducing
the contour to 70\% in some cases.  A few sources in the cluster
core may still suffer confusion (see \S\ref{s:radial}).  The psf
fraction enclosed 
by the region was calculated at five energies, and interpolated for
other energies.  Event lists, spectra, and light curves were extracted
for each source and observation, and response matrices and effective
area files 
(including exposure from multiple CCDs where appropriate) were
constructed for each source.  Background spectra were extracted for
each source using regions sized to include 100 counts and 
excluding mask regions (generally 1.5 times the size of the 90\% psf
contour) around all detected sources.  The effective area functions
were created using CIAO v.~3.02 and CALDB v.~2.26, including the new
CALDB ACIS contamination file to correct for the low-energy quantum
efficiency
degradation\footnote{http://cxc.harvard.edu/ciao/threads/aciscontam/.}.
We then corrected for the finite portion of the flux contained within our 
extraction regions by reducing the effective areas, in an energy-dependent 
manner \citep{Broos02}.
Composite source and scaled background spectra 
were then constructed (separately for the 2000 ACIS-I and 2002 ACIS-S
images), and appropriately weighted response matrices and effective
area files were computed.  The total live exposure time for each
source was also 
recorded, and the spectra were grouped to control the placement of the
first and last bin, thus constricting the energy range for spectral
fitting to 0.5-8.0 keV
for ACIS-I data and 0.35-8.0 keV for ACIS-S data.  

Finally, we computed background-subtracted photometry for each source
in several bands.  We calculate counts and fluxes for bands
used in the globular cluster literature, 0.5-1.5 keV, 1.5-6 keV,
0.5-4.5 keV, 0.5-2.5 keV, and 0.5-6.0 keV \citep[see ][]{Grindlay01a,
Grindlay01b, 
Pooley02a, Pooley02b, Heinke03c}.  We also computed the fluxes for
several other bands for use in making specialized color-color
diagrams; these bands are 0.3-0.8 keV, 0.8-2.0 keV, and 2.0-8.0 keV.  

Below we explain the columns in Tables~\ref{tab:pos}, \ref{tab:pos2000},
and \ref{tab:hr}. All errors are
$1\sigma$, containing 68\% confidence for one parameter of interest.

Column (1): the \Chandra WAVDETECT detection number, following the
detection and naming
convention systems of GHE01a; \citet{Grindlay02, Edmonds03a,Edmonds03b} for
sources detected in the 2000 dataset (up to W184), and adding new
numbers above W184 for new detections. 

Column (2): the IAU-approved source name.

Columns (3,4,5): the position, corrected to the MSP frame; errors in
each coordinate, expressed in seconds (not arcseconds) for right
ascension; and distance from the center of the cluster (taken to be
$\alpha=00^{\rm h}$ 24$^{\rm m}$ 05\fs29, $\delta$=-72\degr 04\arcmin
52\farcs3; \cite{DeMarchi96}). 

Columns (6,7,8): background-subtracted counts in three
standard bands.

Columns (9,10): X-ray luminosities computed for 0.5-6.0 and 0.5-2.5
keV bands.
Photon fluxes were computed from the recorded counts, 
effective apertures, exposure times and effective area functions for
each source.  Assuming a 2 keV VMEKAL spectrum (a reasonable
approximation for the average spectrum of the fainter X-ray sources,
see \S\ref{s:phot} and \S\ref{s:spec} below),
an absorbing column of $1.3\times10^{20}$ cm$^{-2}$, and a distance of 
4.85 kpc, we derive conversion factors of $5.93\times10^{36}$ ergs
photon$^{-1}$ cm$^{2}$ and $4.94\times10^{36}$ ergs photon$^{-1}$
cm$^{2}$ to convert from photon flux to unabsorbed X-ray luminosity in
the 0.5-6.0 keV and 0.5-2.5 
keV energy bands, respectively.  To recover the directly measured
photon fluxes in photons cm$^{-2}$ s$^{-1}$, the reader may divide the 
quoted luminosities by these factors.

Columns (11,12,13): Notes on individual sources: ID type (note that a
? indicates a tentative classification; other IDs are secure
classifications. qLX indicates a qLMXB, AGB indicates an AB from AGB01
identified with an 
X-ray source in this work), other names (AKO 9 from
\cite{Auriere89}, ROSAT X \# from \cite{Verbunt98},
MSP lettering from \cite{Camilo00,Freire01a,Edmonds02b}), and notes
on variability (Y, D and H for 99.9\% 
confidence variability on timescales of years, days, and hours
respectively; D? and H? for 99\% 
confidence variability on day or hour timescales). A c indicates that the
source may suffer from confusion. An m indicates that the source was
identified manually, rather than by WAVDETECT.

Table~\ref{tab:pos2000}: Same as Table~\ref{tab:pos}, except that we
include the psf fraction (column 5) and fractional exposure time
(column 6, compared to 72155 s) instead of the radial
distance. Notes D and H indicate variability seen on day or hour
timescales within the 2000 dataset.  In addition, the notes 
identify (with a ``t'') the three sources detected by WAVDETECT in
2000 that are not detected in the 2002 data; two of these may be
spurious detections.

Table~\ref{tab:hr}: Column (3): Fraction of the 1.5 keV point-spread
function enclosed by our extraction region for each source.  

Column (4): Fractional exposure time for each source, compared to the
nominal exposure time of 281074 seconds.

Columns (5,6,7,8): Counts in four supplementary bands.

Columns (9,10,11,12): Hardness ratios.  HR1, HR2 and HR3 defined as
(H-S)/(H+S), HR4 defined as S/H.  For HR1, H=2.0-8.0 keV, S=0.8-2.0
keV; for HR2, H=0.8-2.0, S=0.3-0.8 keV; for HR3, H=2.0-8.0 keV,
S=0.3-0.8 keV; for HR4, H=1.5-6.0 keV, S=0.5-1.5 keV.

\section{Variability Analysis}\label{s:var}

We here attempt to quantify the variability of our sources on a
variety of timescales; during individual OBS\_IDs (hours), between
OBS\_IDs (days), between the 2000 and 2002 observations (years), and
(for the brightest sources) between the 2002 observations and the
ROSAT observations of the 1990s discussed in \citet{Verbunt98}.  

For intra-OBS\_ID and inter-OBS\_ID variability searches, we exclude
the shortest exposures, which can independently  
detect only a few sources, leaving two exposures in the 2000 data and four
in the 2002 data.  For the 2002 data
we use the 0.3-6.0 keV events, while for the 2000 data we use the
0.5-6.0 keV events. To identify variability on a timescale of hours we applied a
Kolmogorov-Smirnov test to the un-binned arrival times of the events
from each observation\footnote{We used ACIS\_EXTRACT 
v. 3.5 to compute KS probabilities.}.  Sources
that exhibited short-term variability at the 99.9\% confidence level
in at least one observation are marked H for short-term ``hours''
variability in 
Tables~\ref{tab:pos} and \ref{tab:pos2000}, while sources that showed
variability at the 99\% level 
are marked H? for possibly short-term variable.  In this way we
identify two short-term variables (six including possible variables)
in the 2000 dataset, and 25 short-term variables (51 including
possible variables) in the 2002 dataset.  Considering the 146 sources
in the 2000 dataset (and two searched exposures) as 292 trials, three of the
possibly variable sources may not be variable.  Using the four exposures
and 300 sources in the 2002 data, twelve of the possibly variable 
sources may not be variable.  These estimates, based on the KS
probabilities, may overestimate the amount of spurious variability,
since the faintest sources considered do not provide sufficient counts
in a single exposure for meaningful use of the KS test
\citep{Press92}.  All six variable or possibly variable sources in the
2000 dataset show high-confidence short-term variability in the 2002
dataset.

To identify longer-term (timescale of days) variability we compare the
count rates between long exposures of a source (with the same
instrumental setup), ignoring the effect of background (which is 
very low in these observations).  We define a significance level $\sigma$,  

$ \sigma = |(C_j - {\rm (ER)} C_i)/(\sqrt{C_j+{\rm (ER)}^2 C_i })$

\noindent where the exposure ratio ER is the exposure time times the
effective area for 
observation $j$ divided by the exposure time and effective area for
observation $i$ , and $C_j$ and $C_i$ are the counts recorded from the
source in each observation \citep[][]{Feigelson02}.  We identify
sources showing  $\sigma>2.6$ as possibly variable (D?; 99\% confidence),
and sources showing $\sigma>3.3$ as variable (D; 99.9\% confidence) 
between exposures.  Four sources show high-confidence
days-timescale variability between the two 2000 observations, while 31
sources show days-timescale variability (53 including possible
days-timescale 
variability) among the 2002 observations.  Considering the six
possible combinations of 2002 observations, 18 of the possibly
variable sources may not be variable, but again this could be an
overestimate (see above).  

Variability is more complicated to measure between observations using
different instruments. This is because the change in effective areas
at different energies may cause a source, with a spectrum different
from that which is assumed, to appear to be variable even if it is
not.  To reduce this problem (and allow comparison with ROSAT
results for some sources) we compare fluxes in the 0.5-2.5 keV band,
rather than the broader 0.5-6 keV band, and only identify sources with
99.9\% certainty of variability.  We compute $\chi^2$ as

$\chi^2 = (F_{2002}-F_{2000})^2/(\sigma_{2002}^2+\sigma_{2000}^2)$

\noindent using the upper or lower errors appropriately, and identify
variability if $\chi^2 > 10.827$.  These sources are indicated with a
Y, for variability on timescales of years, in Tables \ref{tab:pos},
\ref{tab:pos2000}, and
\ref{tab:hr}.  Three sources in the 2000 data are not seen in the
2002 dataset, including W68 and two other possibly spurious sources.
Some sources in the 2002 data would have been detected in the 2000
data had their emission remained constant, but are not detected in the
2000 data.  To identify these sources we use equation 2 from
\citet{Muno03} to estimate the upper limits (at a signal-to-noise ratio of 3) at the
locations of 2002 sources not detected in the 2000 data.
We use background areas encircling 90\% of the 1.5 keV energy, with radii approximated
(for our limited range of off-axis angles) by
$r=2.65-0.14\theta+0.18\theta^2$, $r$ in pixels and $\theta$ in
arcminutes. Sources with flux in 2002 more than 3 sigma above these upper
limits are identified as clearly variable; we find five such sources. 
 
 We perform the same tests for the nine X-ray sources identified by
 \citet{Verbunt98} in the ROSAT data, taken between  
 1992 April and 1996 November. For these sources we recompute the ROSAT
X-ray luminosities 
using the counts of \citet{Verbunt98}, a 2 keV thermal plasma model,
and PIMMS\footnote{Available at http://asc.harvard.edu/toolkit/pimms.jsp.}.
We approximate the upper limit of detectability to be the flux from X19
(the dimmest source within the core region) for other sources within
the core. For sources outside the core we use the flux from X13 as an
upper limit.  In all cases we identify variability at the 99.9\%
 level.  The results of these tests are listed in
Table~\ref{tab:rosvar}, along with the conclusions from
\citet{Verbunt98} about variability within the ROSAT dataset.

Ascertaining variability is of particular interest because some groups
of X-ray sources in 47 Tuc are expected to show intrinsic variability,
while others are not.  Millisecond pulsars are not expected to show
variability (on the timescales we can detect, using 3.2 second
resolution times), and qLMXBs may not show intrinsic variability if their
X-ray emission is generated by release of heat from their cores
\citep{Brown98}, although they may vary due to eclipsing behavior or
variations in $N_H$ column \citep{Heinke03a}.  On the other hand, CVs
and ABs are expected to show a variety of variable behaviors,
including flickering and flares (due to magnetic reconnection events
in ABs).  The hypothesis that most ABs in the X-ray luminosity range
probed by these observations would appear as transient sources,
flaring to detectability for a few hours and then dropping below our
detection limit, is 
proven incorrect.  Most ABs in our data are detectable in all
observations of equal depth, and many show no evidence of variability
even when $>$100 counts are detected.  

The MSP 47 Tuc-O is identified as long-term variable (3.7$\sigma$) between the
second and fourth 2002 observations, appearing to decrease
continuously in flux from the second to the fourth observation.  The
MSP 47 Tuc-U is also possibly variable within the third 2002
observation, and 47 Tuc-R is possibly variable between observations in
2002.  No
physical mechanism has been proposed, to our knowledge, that would
explain significant variability of the flux from these old MSPs on
these timescales.  We think
it most likely that W39, only a few arcseconds from the core 
of the cluster, is a blend of 47 Tuc-O and another, variable X-ray
source, most likely an AB (see Bogdanov et al.~2005, and
Figure~\ref{fig:inset}).  A portion of 
our sources are certain to be blends, and we have marked some likely
confused sources in Table~\ref{tab:pos}.  47 Tuc-R is only 1\farcs3
from the brighter, variable source W24, which may cause a spurious
detection of variability.  For 47 Tuc-U we can say only that
there is a reasonable probability that one MSP will be spuriously identified
as possibly variable.  The X-ray sources W6, W31, W71, W91,
W96, and W97, identified as potential MSPs by \citet{Edmonds03b}, show
variability or possible variability, and thus should be considered
less likely MSP candidates.  However, three of these (W6, W91, W96) are only
classified as possibly variable. 

\begin{figure}
\includegraphics[scale=.50]{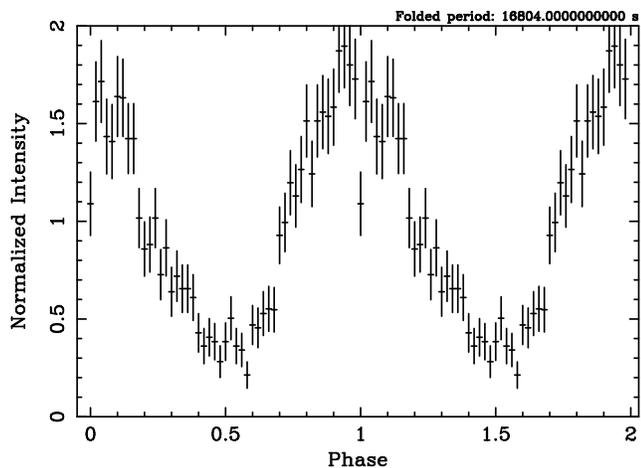}
\caption[X10_lc.eps]{Folded X-ray lightcurve for the CV X10 (W27) on
a 16804 second period.  Two cycles are shown for clarity.  
} \label{fig:X10lc}
\end{figure}

\begin{figure}
\includegraphics[scale=.50]{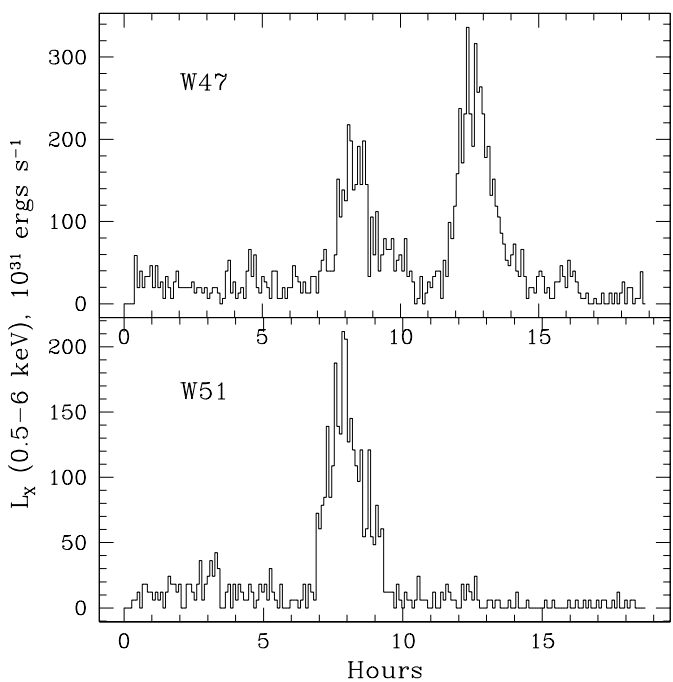}
\caption[W47_W51_lc.eps]{Background-subtracted lightcurves of two
sources showing extraordinary flares. Conversions from countrates to
X-ray luminosities are 
calculated using the time-averaged spectra, which we have confirmed
are reasonable spectral representations of the flares.  These
conversions are $2.2\times10^{34}$ ergs ct$^{-1}$ for W47 (top,
OBS\_ID 2737), an AB, and $2.0\times10^{34}$ ergs ct$^{-1}$ for W51
(bottom, OBS\_ID 2735), a CV.  Each became briefly as 
luminous as any other source in the cluster.  
} \label{fig:lightcurves}
\end{figure}

We note that all sources with more than 1000 counts (13) are identified as
variable, except for the qLMXB X7, the brightest source in the field
($>30000$ counts),
which shows no indication of variability within the 2002 data.
This indicates that the 
mechanism driving the X-ray emission is probably different from that in Aquila
X-1 and Cen X-4, where significant variability on short time scales
is generally thought to be due to accretion 
\citep[cf.][]{Rutledge02b, Campana03, Heinke03a}.  
X10 displays very strong periodic variability (P=16804 s),
with a phased light curve (Fig.~\ref{fig:X10lc})
similar to those seen in the magnetic CVs known as polars
\citep{Ramsay04}.  The AB W47 and the CV
W51 each show spectacular flaring 
behavior in the 2002 dataset, reaching an energy output of
several $10^{33}$ ergs s$^{-1}$ at the peak (Fig.~\ref{fig:lightcurves}).  
Searches for possible periodicities, long-term
variability between the 2000 and 2002 \Chandra observations (and ROSAT 
observations), and 
more detailed characterization of timing results, will be presented in
future papers (Grindlay et al.~2004, in prep.). 

\section{Radial Distributions}\label{s:radial}

\begin{figure}
\includegraphics[scale=.50]{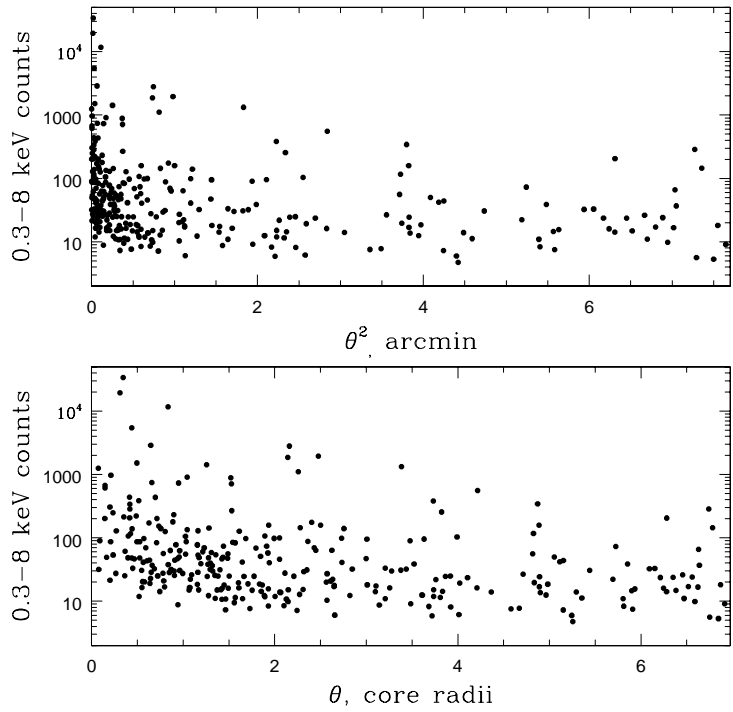}
\caption[sm_rad_prof2.eps]{Distributions of extracted source counts
vs.\ radial distance from center of 47 Tuc \citep{DeMarchi96}.  Top:
Radial distance plotted in units of arcmin$^2$, so background sources
should be evenly distributed.  Bottom: Radial distance plotted in
units of 47 Tuc core radii (1 $r_c$=24\arcsec).  Beyond about 100\arcsec
(2.77 arcmin$^2$, 4.17 $r_c$), the source density appears flat.
} \label{fig:radprof}
\end{figure}

In order to understand how many of the sources we have detected are
background sources vs.~cluster members, we study the 2002 detected sources' 
radial distributions. (The 2000 data is significantly incomplete 
due to chip gaps, so we disregard it in this section.)  The effective
area of \Chandra$\!$'s mirrors does not 
decrease significantly within our 2\farcm79 (47 Tuc half-mass radius)
survey area (effective area at 2\farcm79 is $\sim$98\% of on-axis value
for 1.5 keV photons, Chandra Proposer's Observatory Guide), and our
covered area lies almost entirely (99\%) on the ACIS-S3 chip.  Thus
our sensitivity does not change dramatically across our surveyed area.  In
Figure~\ref{fig:radprof}, we show the distribution of 0.3-8 keV
counts vs.~distance from 
the cluster center, both in core radii (24\arcsec$\!$) and in units of
square arcminutes, where background sources should be evenly spread. 
Figure~\ref{fig:radprof} and careful examination of the data indicate
that essentially all
sources above 20 counts (0.3-8 keV) have been detected (though in many 
cases we detect much weaker sources).  

This is not strictly true where 
sources lie close enough together to overlap; the existence of 75
sources above 20 counts within the core of 47 Tuc produces a 3.5\%
chance of one of these sources  
landing within 0\farcs5 of another source.  This suggests that 2 or
3 sources above our detection limit within 47 Tuc's core are
{\it unresolvably} confused with other sources (since the pixel scale
of the ACIS detector is only 0\farcs492).  A further $\sim6-7$ should be 
located within 1\arcsec\ of other sources, which may be recognized as
confused  
sources inconsistent with the \Chandra point-spread function.  Many
of these should be separable using more complicated 
detection algorithms, such as two-dimensional Kolmogorov-Smirnov tests 
\citep{Metchev02}; this work is in progress.  An example 
of two sources that are confused in our source detection scheme is 47 
Tuc-S and 47 Tuc-F, two radio MSPs located 0\farcs74 from each other and
detected as the single, clearly confused source W77.  
An additional 2-3 faint sources may also be lost within 2-3\arcsec\ of
the three brightest sources in  
47 Tuc, X7, X9 and X5, where the point-spread function wings are
capable of obscuring sources at our detection limit.  Thus, we
probably miss some 12 sources from inside the core radius.  
We bear in mind these caveats in the analysis that follows. 

We first estimate the background density.  It can be seen
directly from Figure~\ref{fig:radprof} that the number of sources per square
arcminute appears flat beyond roughly 100\arcsec\ (2.78 arcmin$^2$).
A background level of 1.5$\pm0.3$ sources arcminute$^{-2}$ (error
derived from Poisson statistics) gives an
asymptotically flat cumulative excess of sources above the background
beyond $\sim$100\arcsec, for a total of 37 background sources within
the half-mass radius.  
Therefore, the 24 sources with $>20$ net counts
that lie between 100\arcsec\ and the 47 Tuc half-mass radius are
consistent with being background sources.  

We estimate the expected background extragalactic source numbers from
the cumulative number counts (0.5-2.0 keV band, where we detect nearly 
all of our sources) of \citet{Brandt01},
\begin{equation}
N(>S)=3970(S/10^{-16} {\rm \,ergs\, cm}^{-2}\, {\rm
  s}^{-1})^{-0.67\pm0.14}{\rm \,deg}^{-2}  
\end{equation}
which was derived
from sources at the flux levels to which we are sensitive. Assuming a
$\Gamma=1.4$ photon index power-law spectrum \citep{Giacconi01,Brandt01}, and
$N_H=1.3\times10^{20}$ cm$^{-2}$, 20 counts is a 0.5-2 keV flux of
$1.63\times10^{-16}$ ergs cm$^{-2}$ s$^{-1}$, giving 2860$\pm200$
sources degree$^{-2}$.  Thus, we should find $19\pm4$ background AGN
 above 20 counts within the half-mass radius.  This is roughly half of
 the number of sources we detect, which 
suggests an enhancement over the deep-field number counts of
\citet{Brandt01}.  This level of enhancement is unlikely to be due to
cosmic variance in extragalactic number counts \citep{Kim04}.  Below
20 counts, uncertainties on our flux 
measurements due to noise and confusion may cause a preferential shift
of our fluxes to higher values \citep[a version of Eddington bias,
][]{Murdoch73, Muno03}, but this should not be a large problem above
20 counts since the slope of our overall luminosity function is
relatively flat (see \S\ref{s:lf}).
 This enhancement could be caused by background X-ray sources in the nearby 
Small Magellanic Cloud.  Alternatively, these X-ray sources may reside
in 47 Tuc's halo, either because they are relatively low-mass binaries
pushed out by mass segregation, or because they were 
 generated by primordial binaries in the distant halo
 \citep{Davies97,Gendre03a}.  Some close binaries are known to exist
 in 47 Tuc's halo \citep{Kaluzny98,Weldrake04}, and blue stragglers in
 47 Tuc's halo are thought to be generated by primordial binaries
 \citep{Ferraro04}.  The nature of this X-ray background population
 will be probed in future papers that examine the full \Chandra fields.

\begin{figure}
\includegraphics[scale=.50]{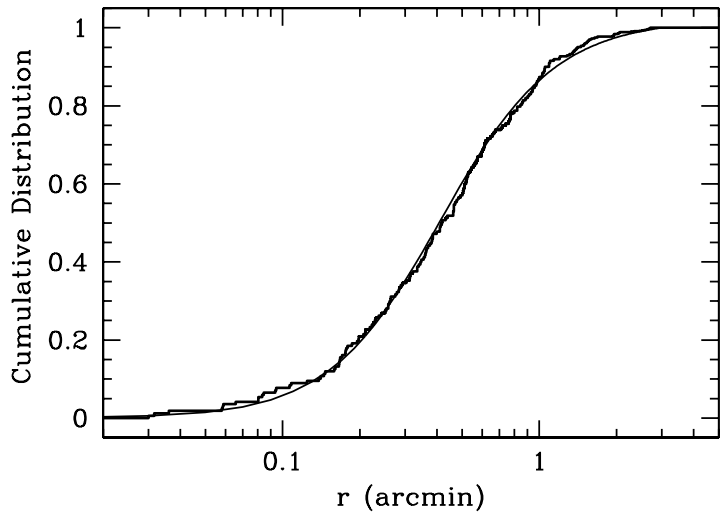}
\caption[Radial_cum.eps]{Plot of cumulative radial distribution of
47 Tuc X-ray sources (above 20 counts, 0.3-8 keV energy band) fit to
an analytical model (Eq.~2) for their radial distribution. The best
fit, plotted here, gives a ratio of  
average source mass to typical stellar mass of q=1.63.} \label{fig:radcum}
\end{figure}

The radial distribution of objects in a dynamically relaxed cluster
allows an estimate of the average mass of those objects, in terms of a
mass ratio between the object and the average mass of the typical
stars that define the central gravitational potential
\citep{Grindlay84, Grindlay02}. \citet{Heinke03d} describe a procedure for
estimating the typical qLMXB mass from the spatial distribution of a
sample of 20 probable qLMXBs in seven clusters.  This procedure is
based on maximum-likelihood fitting of a parametrized form to the
radial profile of the source distribution.  The key parameter is the
ratio $q=M_X/M_*$ of the source mass to the mass of the typical stars
that define the optical core radius.  This approach assumes that the
spatial distribution of these typical stars is well described by a
classical \citet{King66} model, reasonable for 47 Tuc
\citep{Howell00}.  Then the radial profile for the source density
takes the form 
\begin{equation}\label{profile_fcn}
S(r) = S_0 \, \left[1 + \left({r \over r_{c\ast}}\right)^2
\right]^{(1-3q)/2},
\end{equation}
where $S_0$ is an overall normalization and $r_{c\ast}$ is the optical
core radius determined for turnoff-mass stars (here we use 24\arcsec$\!$).
We correct the $>20$ count source sample for background contamination
using the 
estimate of 1.5 sources arcmin$^{-2}$, and the bootstrap resampling
procedure described in \citet{Grindlay02}.  The best-fitting value for 
$q=M_X/M_*$ is $1.63\pm0.11$, consistent with the value of
$q=1.5\pm0.25$ found for soft sources in \citet{Grindlay02}, and the
cumulative radial distribution is compared with this model in
Figure~\ref{fig:radcum}.  Restricting our sample to sources with $>40$
counts, we find 
identical results, indicating that crowding in the core does not
dramatically affect these results.  

To estimate the mass of these sources requires an estimate of the mass
of the core stars to which we are comparing our sources.  The core
radius we use is from \citet{Howell00}, which uses a limiting $U$
magnitude of 18.11, or $V$=17.6.  Using the model isochrones of
\citet{Bergbusch01} as displayed in \citet{Briley04}, the mass range
included in \citet{Howell00} is from 0.865 \Msun to 0.915 \Msun.  
We take the average mass of the stars which determine the core radius
of 47 Tuc to be $0.88\pm0.05$ \Msun, allowing for some uncertainty in
the modeling (e.g., in the distance to 47 Tuc).  We also include the
8\% uncertainty in Howell et al.'s core radius determination.  
Our result for the X-ray sources is an average mass of
$M_X=1.43\pm0.17$ \Msun\ for the X-ray sources.  This is
consistent with our expectations for a mixture of qLMXBs, CVs, ABs,
and MSPs, all of which are 
significantly heavier than single main-sequence stars.  

We can also constrain the masses of individual populations within the
GO-8267 \HST\ field, incorporating the radial incompleteness of the
WFPC2 field.  Including all 60 ABs and candidate ABs within this field
produces $q=1.12\pm0.10$, implying a typical mass $M_X=0.99\pm0.13$
\Msun.  This is significantly lower (at the $4.6\sigma$ level) than the
average q value of the X-ray sources as a whole, calculated
above.  This does not take into account our reduced
sensitivity to X-ray sources in the cluster core.  Restricting our
fits to the 43 ABs with $>20$ counts gives $q=1.32\pm0.13$, only
$2.4\sigma$ from the mean X-ray source $q$.  However, there is a
correlation between X-ray flux and AB mass \citep{Edmonds03b}, so
that ABs brighter in X-rays may be expected to be more massive.  Starting from
the full sample of 60 ABs and candidates within the GO-8267 field of view,
we exclude an inner core of increasing size, from 0 to 1.5 core
radii.  The value of $q$ produced by each sample remains essentially
the same (ranging from 1.10 to 1.14), leading us to conclude that the
relatively low mass of ABs (compared to the average X-ray source) is
robust. 

We can perform the same test (with lesser significance) for the CV and
MSP populations.  For the 22 CVs and CV candidates within the GO-8267
field, we derive $q=1.49\pm0.20$, inferring a mass of $M_X=1.31\pm0.22$
\Msun, consistent with our rough expectations for CVs containing heavy
white dwarfs plus secondary stars.  
For the 22 MSPs
and MSP candidates (not including R) in the GO-8267 field of view, we find
$q=1.42\pm0.16$, implying $M_X=1.25\pm0.19$ \Msun.  This group seems
less massive than  
expected, which suggests that several of the MSP candidates 
identified as possible MSPs by \citet{Edmonds03b} are not actually MSPs  
\citep[as suspected by][]{Edmonds03b}.  

For the 17 securely identified MSPs (not including R) with previously 
known positions \citep[radially complete to 6\arcmin,][]{Camilo00}, we
derive $q=1.67\pm0.14$, implying $M_X=1.47\pm0.19$ \Msun.  
We can compare this number with the expected average mass of the known MSP
binary systems in 47 Tuc, using the total mass of 10 binary companions
\citep[1.33$\pm0.11$ \Msun, using a mean value of 1/sin~$i$ of
  $4/\pi$,][]{Chan50,Backer98}  to find an 
average neutron star mass of $1.39\pm0.19$ \Msun.  Comparing this with
the average pulsar mass of  1.35$\pm0.05$ \Msun\ \citep{Thorsett99}, we
see that MSPs in 47 Tuc have not accreted a substantial 
amount of mass (less than 0.23 \Msun) during their recycling
\citep[in agreement with][]{Thorsett99}.

\section{Photometry}\label{s:phot}

\begin{figure}
\includegraphics[scale=.80]{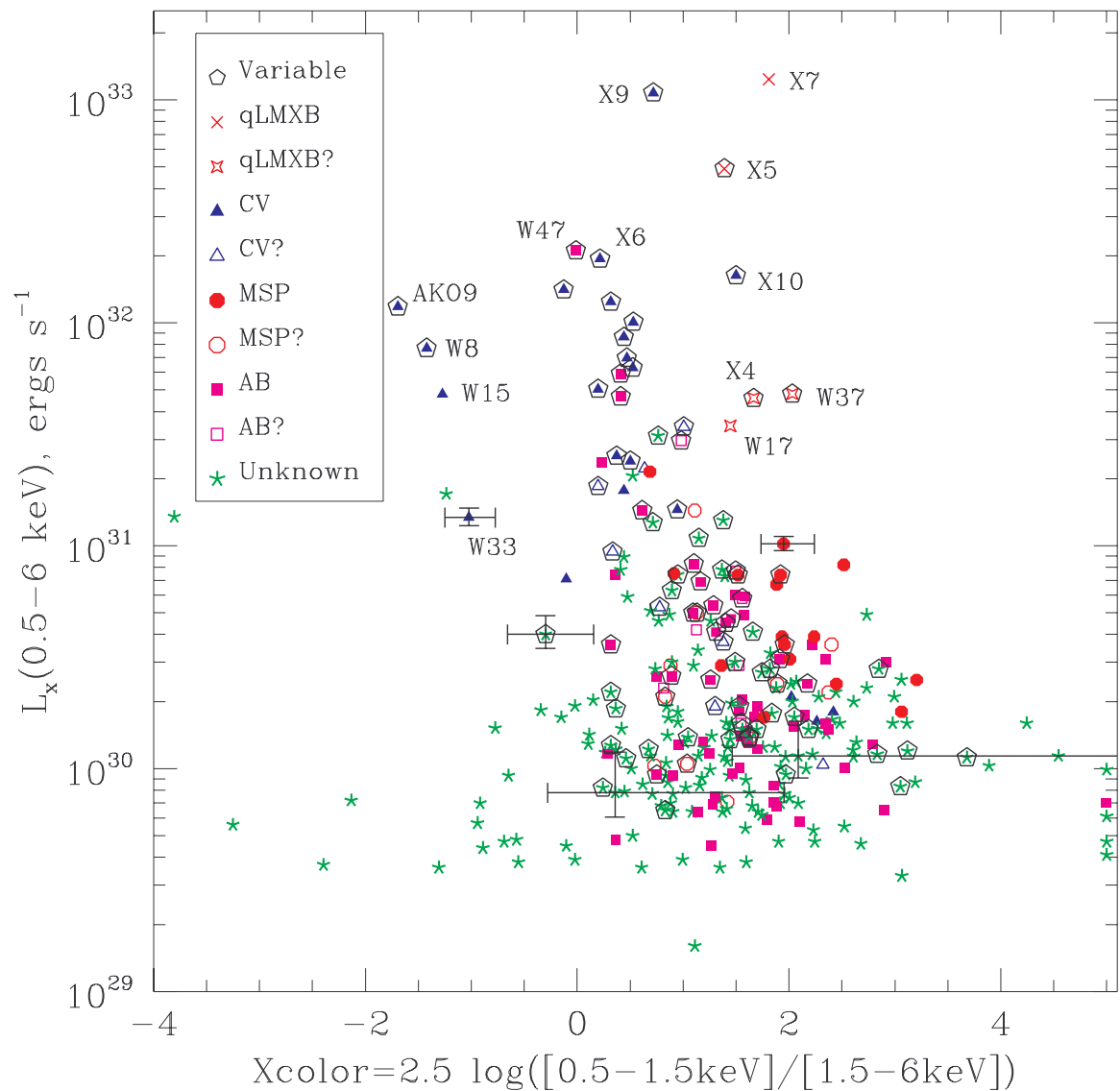}
\caption[XCMD_april_CMYK.eps]{X-ray CMD, plotting X-ray luminosity
  (0.5-6 keV, absorbed) against  
hardness (increasing to the left) for 47 Tuc X-ray sources from 2002
data.  X-ray luminosities are taken from 
spectral fits to thermal plasmas (Table~\ref{tab:spec}, column 7), or
for fainter 
sources, from the photon fluxes using a conversion assuming a 2 keV
VMEKAL spectrum (Table~\ref{tab:pos}).  Since the   Variable sources ($>$99\%
confidence) are indicated with pentagons.  Only a few error bars are
plotted, to improve readability.  Several sources of particular
interest are indicated with their most common names.}\label{fig:XCMD}
\end{figure}

We produced an X-ray version of a color-magnitude diagram for the 2002 
data (Fig.~\ref{fig:XCMD}), plotting broad-band luminosity vs. the
hardness ratio $2.5\times$log[(0.5-1.5 keV cts)/(1.5-6 keV cts)]
\citep[GHE01a,][]{Grindlay01b,Pooley02a,Pooley02b}.  We found the
most useful quantity for the y-axis to be the observed 0.5-6 keV flux,
which we took from spectral fits to a VMEKAL model
(\S\ref{s:spec} below) when sufficient counts were available for
simple spectral fits. (We use the observed 0.5-6 keV flux, not
correcting for absorption, since in many cases the $N_H$ column is
poorly determined.)  
Otherwise we used the X-ray luminosities computed using a 2 keV
VMEKAL model (see \S\ref{s:spec}), and the photon fluxes computed above
(\S\ref{s:extract}).   (The actual luminosities will be dependent upon 
the intrinsic spectral shape.  We choose a 2 keV VMEKAL model as a reasonable
average for these spectra, since the hardness ratios are poorly
constrained for these faint sources.)
  We have indicated all the major classifications of identified 
sources, and whether each source shows variability above the 99\%
confidence level within the 2002 data, or shows variability between
the 2000 and 2002 datasets.  This plot is qualitatively similar to the
X-ray CMD shown in GHE01a for the 2000 data, but our choice of y-axes
shifts the harder sources up on the diagram. 
 
\begin{figure}
\includegraphics[scale=.80]{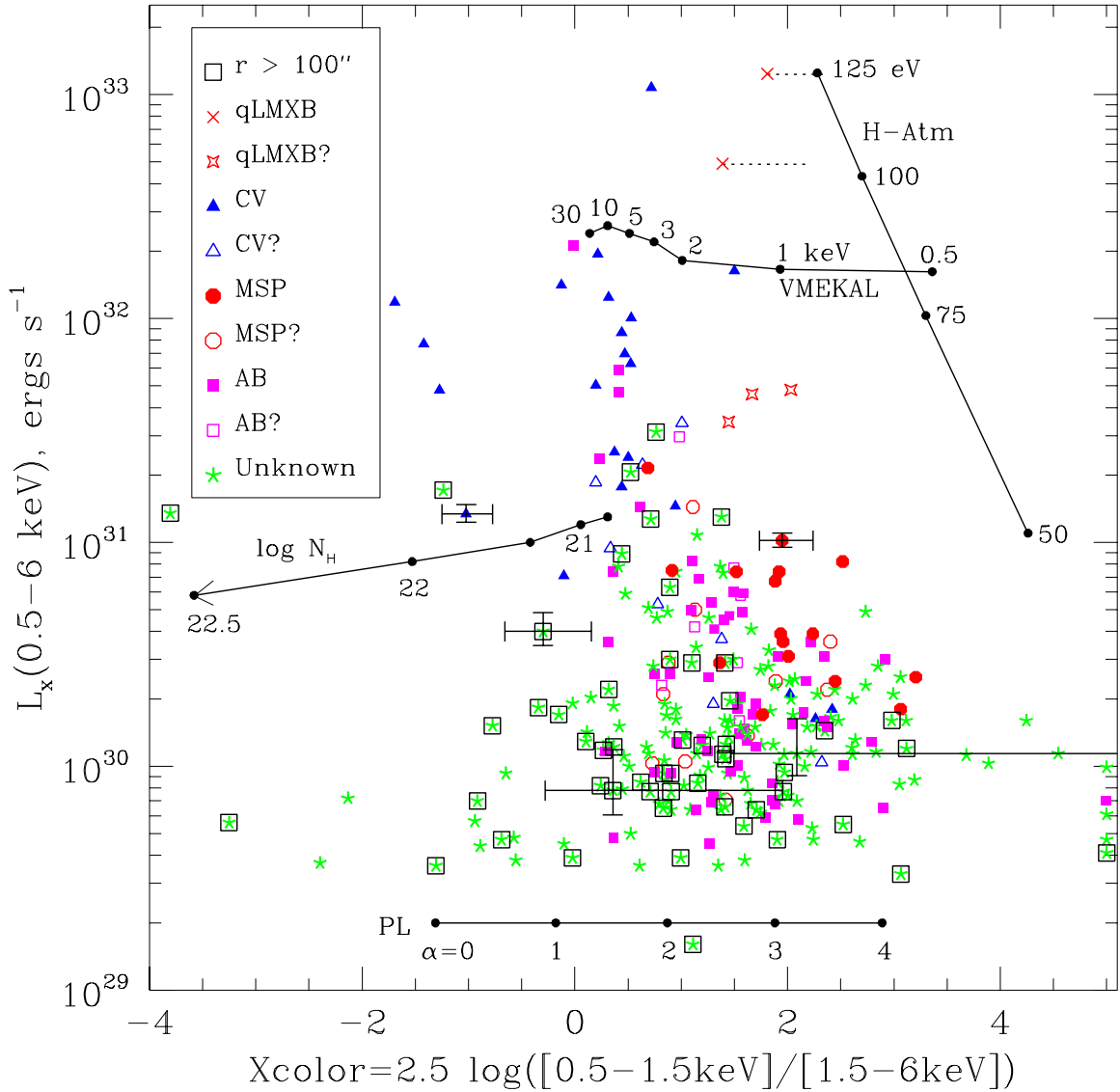}
\caption[XCMD_lines_CMYK.eps]{X-ray CMD, plotting X-ray luminosity (0.5-6
  keV) against  
hardness (increasing to the left) for 47 Tuc X-ray sources from 2002
data, as in Figure~\ref{fig:XCMD}. Sources located beyond 100\arcsec\
  are indicated with squares, and are largely background.
  The location of several 
model spectra are also indicated, assuming the cluster $N_H$ column
except where indicated. The spectra are marked with units of
  temperature (for H-atm in eV and VMEKAL in keV), photon index (for
  the power-law), and $N_H$ in log cm$^{-2}$.  The vertical locations of the 
VMEKAL, PL, and $N_H$ lines are arbitrary, while the effect of increasing 
$N_H$ (shown here for a 10 keV VMEKAL spectrum) varies depending on the 
input spectrum.  Dotted lines next to the qLMXBs X5 and X7 indicate the 
shift in their Xcolors when only the subarray data (with mitigated
  pileup) is used.  
}\label{fig:lines}
\end{figure}

We have plotted this X-ray color-magnitude diagram again in
Figure~\ref{fig:lines}, indicating X-ray sources located at distances
beyond 100\arcsec  
by placing squares around the symbols (all but one of unknown type).  These
sources appear to be mostly background sources, and many are likely
AGN.   At least 64\% of the background objects in the figure should lie
beyond 100\arcsec.  We have
also plotted the positions of various spectral models, with the known
cluster absorption unless otherwise indicated. 
The equivalent hardness ratios from 
power-law spectra  are indicated at the bottom, for several 
representative photon indices.   Hydrogen-atmosphere neutron star models
are plotted for a range of temperatures (in eV), assuming a 10 km
radius (adding a contribution from a 
power-law spectral component would harden these spectra).  Thermal
plasma models (VMEKAL in 
XSPEC), using 47 Tuc elemental abundances, are plotted for a constant
volume emission measure (VEM) of $2\times10^{55}$ cm$^{-3}$, for a range of
temperatures in keV (indicated).  The effect of increasing $N_H$ upon
a 10 keV thermal plasma spectrum (with a lower, arbitrary VEM) is
indicated, for several values of $N_H$. 

It can be clearly seen that most CVs and ABs follow the same trends in 
these diagrams.  Above $L_X$(0.5-6 keV)$=10^{31}$ ergs s$^{-1}$, most are
consistent with thermal plasma of 5-10+ keV. Below $L_X=10^{31}$ ergs
s$^{-1}$, their temperatures tend to decrease, but with increasing
scatter (the scatter in the sources is significantly larger than the
scatter due to counting statistics).  For their hardness ratios, MSPs
appear to be slightly 
brighter on average than most other source classes.  Several sources
are significantly harder than a thermal plasma spectrum can become
without added absorption.  Three of these sources are the known
eclipsing CVs AKO9, W8, and W15
\citep[GHE01a,][]{Edmonds03a,Knigge03}.  The fainter hard sources are
predominantly located beyond 100\arcsec\ from the cluster, indicating
that many are background AGN with intrinsic absorption \citep[as
suggested by][]{Grindlay02}.  

Several bright soft sources can be identified, which do not follow our 
expectations.  The bright and extremely soft CV X10 is discussed in
\S\ref{s:var} and \S\ref{s:spec} and is a likely magnetic CV showing a soft
excess at its accreting magnetic pole(s).  Few sources fall near the
canonical H-atmosphere qLMXB track, which is well-populated in some
other globular clusters \citep{Heinke03d}.  For the known qLMXB X7
this can be explained by the effects of photon 
pileup in the detector artificially hardening the spectrum. For the
eclipsing qLMXB X5, this is due to a large and highly 
variable gas column as well as pileup hardening.  
We have indicated (with dotted lines) the shifts in
X-ray color for the known qLMXBs X7 and X5, by computing their X-ray
color during subarray exposures, where the effects of pileup are
decreased (for X5, we use only OBS\_ID 3385, in
which X5 is brightest).  We also identify three
unusual, moderately bright yet soft sources (W37, X4 or W125, and
W17).  These sources could be qLMXBs if they have very large gas
columns or significant 
nonthermal spectral components.  We address these sources further in
\S\ref{s:spec}, and in \citet{Heinke04b}.  

\begin{figure}
\includegraphics[scale=.80]{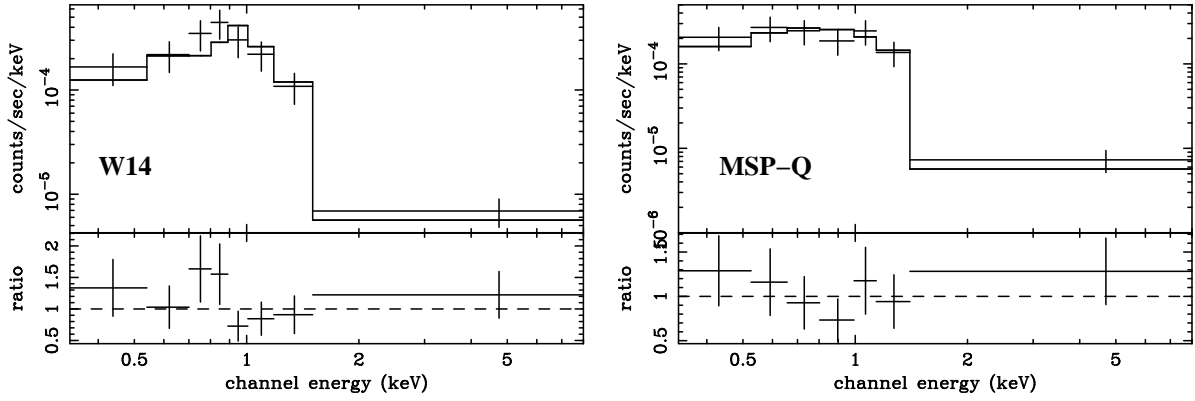}
\caption[ABspec.eps]{Energy spectra of the AB W14 (left) and the MSP
47 Tuc-Q (W104, right).  W14 is fitted with a thermal plasma spectrum
with temperature $1.0\pm0.5$ keV and $N_H=1.3\times10^{20}$ cm$^{-2}$, 
and shows prominent Fe-L emission lines (included in this fit).  The
MSP 47 Tuc-Q is poorly 
fit by a thermal plasma spectrum, and is here fitted by a 137 eV 
hydrogen-atmosphere model spectrum (details in Bogdanov et al.~2005). 
 }\label{fig:ABspec}
\end{figure}

We constructed color-color plots, using different hardness ratios of
the form (H-S)/(H+S), with the bands 2.0-8.0 keV=H, 0.8-2.0 keV=S
(HR1), 0.8-2.0 keV=H, 0.3-0.8 keV=S (HR2) and 2.0-8.0 keV=H, 0.3-0.8
keV=S (HR3).  We select these bands in order to maximize the
differences between a thermal plasma model and emission in the form of 
a power-law.  At low temperatures thermal plasma (even at the low
metallicity of 47 Tuc) will show a relatively larger flux in the 0.8-2.0
keV band due to Fe L-shell line emission (see Fig.~\ref{fig:ABspec}, 
showing the spectra of a typical faint AB and an MSP).  Since we
expect the spectra of faint, soft ABs and CVs to be generated by hot
optically thin plasma, this 
allows some differentiation between them and MSPs, which should not
show Fe L-shell line emission.  

\begin{figure}
\includegraphics[scale=.80]{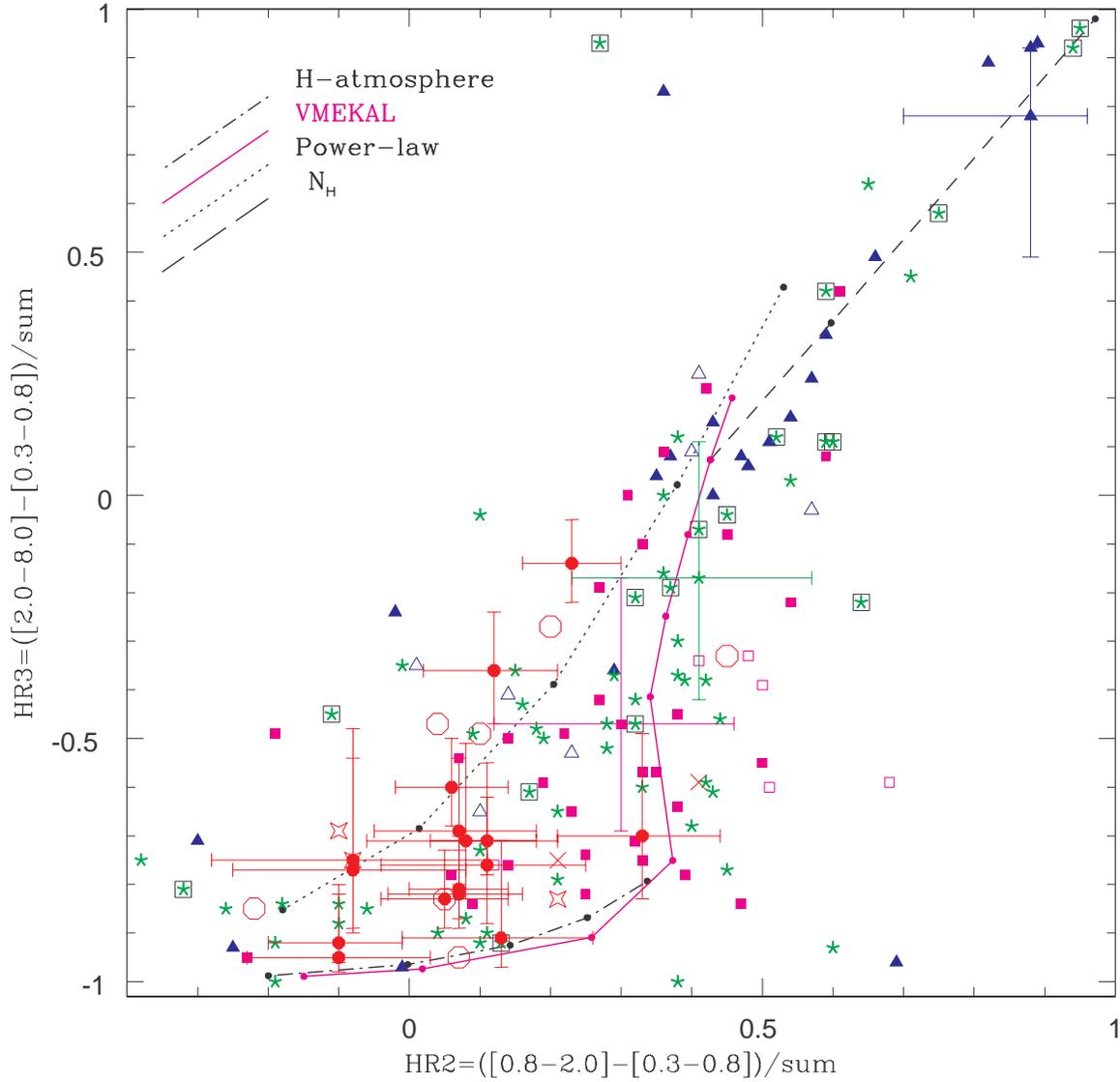}
\caption[sm_color_CMYK.eps]{Color-color diagram for 47 Tuc
sources with more than 30 counts.  Symbols as in Figure~\ref{fig:lines}.
Error bars are plotted for all MSPs, and a few representative
faint sources.  Model 
spectra tracks are plotted, with dots representing the following
values from lower left to upper right: H-atmosphere for 75, 100, 125,
150, 175 eV; VMEKAL thermal plasma model (see text) for 0.4, 0.5, 0.7,
1, 2, 3, 5, 10, and 30 keV; 
power-law, photon index $\alpha$=3, 2.5, 2, 1.5, 1; and effect of
increasing $N_H$ upon a 10 keV VMEKAL spectrum for $1.3\times10^{20}$
(standard), $10^{21}$, and $10^{22}$ cm$^{-2}$. 
}  \label{fig:colorcolor}
\end{figure}

In Figure~\ref{fig:colorcolor} we plot HR2 against
HR3 for the subset of sources with more than 30 total counts 
(including all known MSPs).  We exclude faint sources in order to
decrease the scatter.  Source symbols are the same as in
Figure~\ref{fig:lines}. Errors are  
indicated for all MSPs, and a few other relatively faint
representative sources.  A clear  
difference can be seen in the distributions of ABs and 
CVs in the diagram vs.\ the distribution of MSPs. We have overplotted the
locations of 
three spectral models for a range of parameters: a thermal plasma
model with 47 Tuc abundances (VMEKAL), a hydrogen-atmosphere neutron
star model (since the normalization is not relevant, this is
appropriate for the surface of either MSPs or qLMXBs), and a power-law
with specified photon  
indices.  We have also indicated the effect of increasing $N_H$ for a
10 keV thermal plasma model. 

It is clear from Figure~\ref{fig:colorcolor} that a thermal plasma model, with
temperatures generally between 0.7 and 10 keV, is a
reasonable description of the colors of most of the ABs and CVs (some
with increased $N_H$), and a 
large number of the unknown sources.  The MSPs, on the other hand, lie 
between the tracks for hydrogen-atmosphere neutron star models and
power-law models.  
\citet{Grindlay02} used a similar color-color method on the 2000 data alone to
indicate that the MSPs in 47 Tuc were dominated by thermal emission.
Here we see that a two-component spectrum is likely to best explain
the overall data, including a hydrogen-atmosphere model between 75 and
175 eV and a power-law component with photon index between 3 and 1.5.
The MSP spectra and colors are analyzed in detail in \citet{Bogdanov04}.  

It appears that the unknown 
sources have X-ray colors similar to the identified
sources. 
We attempt to constrain the relative fractions of the unidentified
sources by comparison with the colors of known
sources in \S\ref{s:disc}.

\section{Luminosity functions}\label{s:lf}

\begin{figure}
\includegraphics[scale=.80]{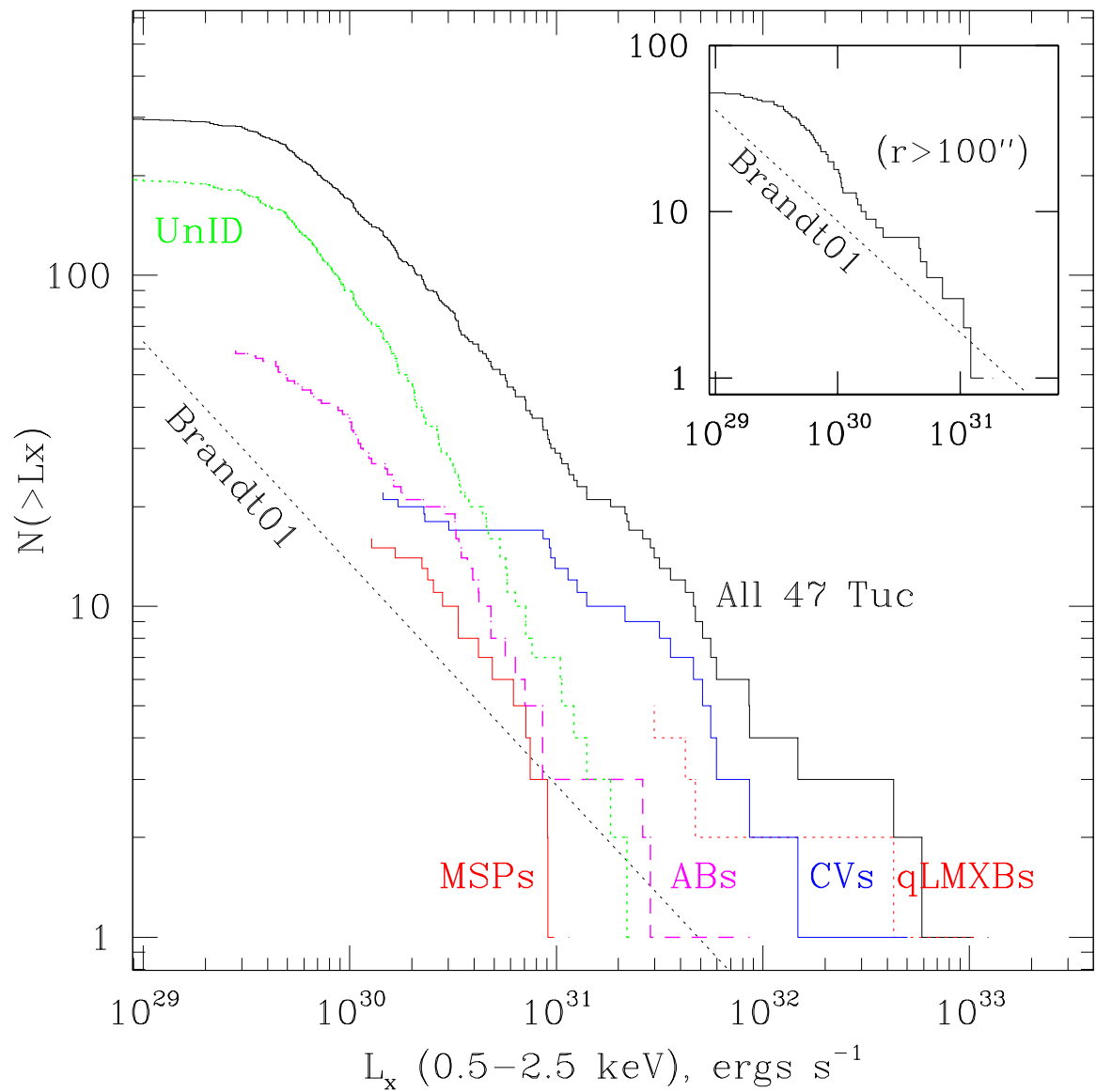}
\caption[CumLx_plot2_bw.eps]{Cumulative luminosity functions for the
different classes of identified sources in 47 Tuc: CVs, ABs,
qLMXBs (including W37, W17 and X4/W125), and MSPs.  Luminosity functions are
also plotted for the total source population, the unknown sources
(including candidate ABs, MSPs, and CVs), and the predicted
extragalactic source counts from \citet{Brandt01}. 
We begin to be incomplete below 20 counts, roughly $8\times10^{29}$
ergs s$^{-1}$. The inset shows sources beyond 
100\arcsec, which have a radial distribution consistent with being
background, but exceed the \citet{Brandt01} prediction (shown for that 
area). (Please see the electronic edition of the Journal
for a color version of this figure.)
} \label{fig:CumLx}
\end{figure}

\begin{figure}
\includegraphics[scale=.80]{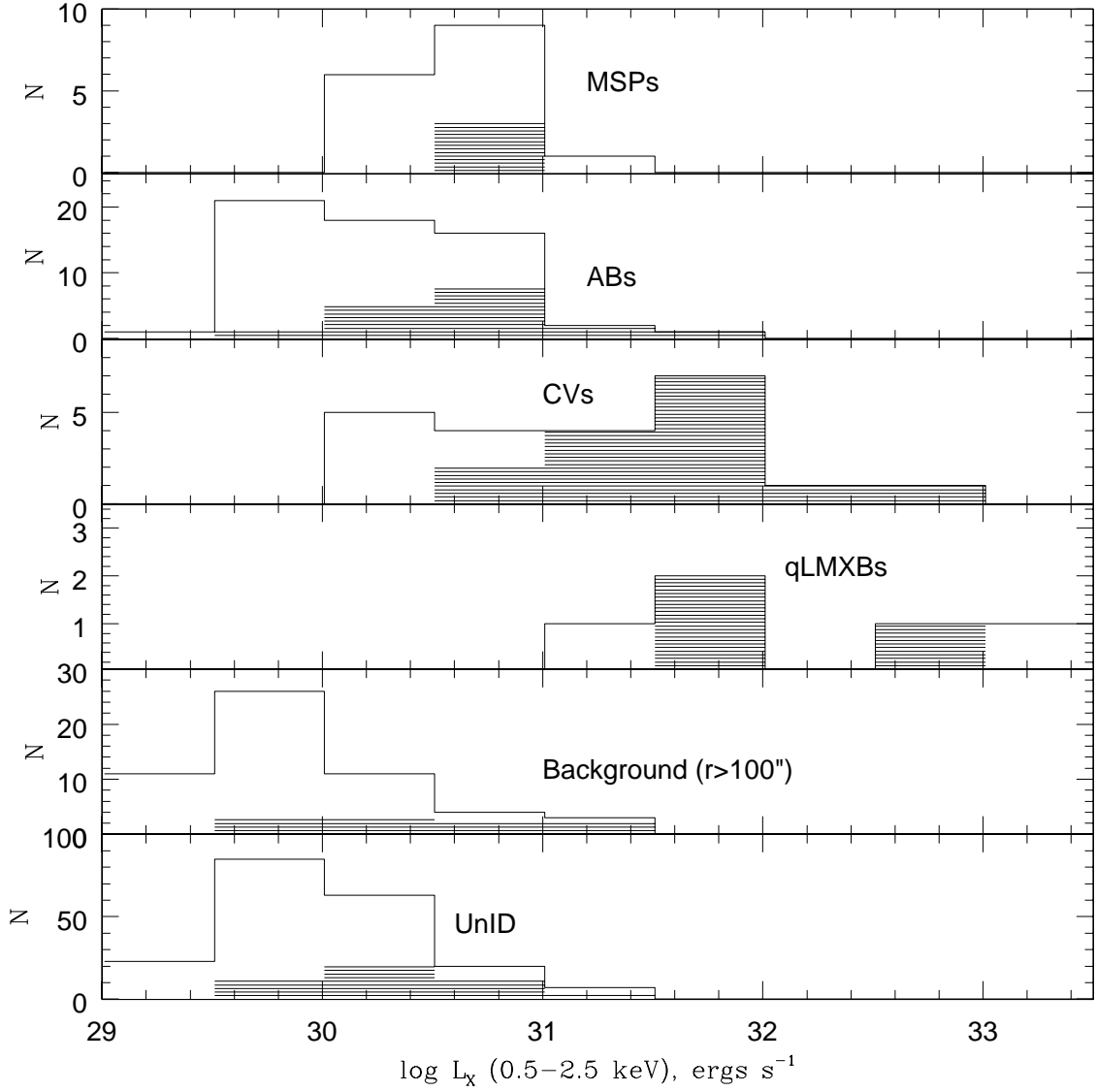}
\caption[varhist.eps]{Differential number counts for the
different classes of identified sources in 47 Tuc: CVs, ABs,
MSPs, and qLMXBs (including W37, W17 and W125), as well as putative
background sources and unknown
sources (not including sources beyond 100\arcsec). The shaded
histograms indicate those sources that were detected to be variable at 
$>99$\% confidence (see text).   Our completeness limit is 
$\sim8\times10^{29}$ ergs s$^{-1}$, but over half the sources are detected
below that limit.  
} \label{fig:DiffLx}
\end{figure}

We construct luminosity functions for the X-ray sources seen in the
2002 data using the 0.5-2.5 keV fluxes from Table~\ref{tab:pos}.   
We show the cumulative luminosity functions of each source class in
Figure~\ref{fig:CumLx}, and the differential number counts in
Figure~\ref{fig:DiffLx}.   We
also plot the expected background counts, using the extragalactic
source density from \citet{Brandt01}.  Our putative completeness limit 
is $L_X$ (0.5-2.5 keV)$=8\times10^{29}$ ergs s$^{-1}$, although many
sources are detected below this limit.  The inset of
Figure~\ref{fig:CumLx} illustrates the
luminosity function of the apparent background 
sources, which is substantially above the expected extragalactic
counts.  Significant cosmic variance is not seen in extragalactic
\Chandra source counts \citep{Kim04}, so another population of objects
is indicated, 
which may be halo sources in 47 Tuc \citep{Davies97}.  If so, the
resemblance of their luminosity function to ABs suggests they are
primordial short-period ABs, and represent the progenitor population
to the halo blue stragglers \citep{Ferraro04}. 
 Figure~\ref{fig:DiffLx} also shows the sources in each luminosity bin 
which we identify as variable or possibly variable.  

The forms of the MSP and qLMXB differential luminosity functions
do not appear to resemble power-laws.  The MSPs were not X-ray
selected, and qLMXBs have not been observed at luminosities below those of our
faintest candidate qLMXBs.  We describe the luminosity functions of these two
groups with lognormal distributions 
rather than power-laws, while we fit the total source numbers,
background sources, CVs and ABs with power-law distributions, using the
maximum-likelihood formalism of \citet{Crawford70}.  
To avoid
Eddington bias strongly affecting the low ends of our luminosity
functions, we fit the luminosity 
functions only for sources with $\gsim$25 counts, or $10^{30}$
ergs s$^{-1}$ for our data \citep{Murdoch73}.  

For the total source population of 47 Tuc, we find a good power-law
fit, $N(>S)\propto S^{-\alpha}$, with $\alpha=0.71\pm0.05$, consistent
with the fit reported by \citet{Pooley02b}, $\alpha=0.78^{+.16}_{-.17}$.  
For the AB population, our fit gives $\alpha=0.88\pm0.14$. We note
that the median X-ray luminosity for BY Draconis 
systems in the field is $1.6\times10^{29}$ ergs s$^{-1}$
\citep{Dempsey97}, so we anticipate many more ABs in 47 Tuc below our
detection limit.  
For the background sources, the calculated index is
0.94$\pm0.22$.  This is significantly steeper than the extragalactic
luminosity function of \citet[][see Eq.\ 1]{Brandt01}, as can be seen in
Figure~\ref{fig:CumLx}.

The CV luminosities can be described by a power-law with
$\alpha=0.31\pm0.04$. The CVs can also be described by a lognormal
distribution with mean log$L_X$(0.5-2.5 keV)=31.2 and standard
deviation $\sigma_{\rm log Lx, CV}$=0.67.  
This is substantially brighter than the average X-ray luminosities of
galactic CVs, as discussed in \citet{Edmonds03b}.  However, since the
known CVs are identified only from the (relatively bright) X-ray
sources in GHE01a, we do not 
yet know the true distribution of CV X-ray luminosities in 47 Tuc. 

The (five) qLMXB (and candidate) X-ray luminosities  
have a mean log$L_X$(0.5-2.5 keV)=32.1 and standard
deviation $\sigma_{\rm log Lx, qLMXB}$=0.7.
The MSP luminosities have 
a mean log$L_X$(0.5-2.5 keV)=30.6 and standard
deviation $\sigma_{\rm log Lx, MSP}$=0.28 (derived without the
unresolved pairs F/S and G/I, which are consistent with this distribution).  
%Check for ABs, mean=30.14, sigma=0.51.
MSPs in the field often have rather uncertain distances.  This may be the
best test of the luminosity distribution of old MSPs, since the
distance and reddening to these MSPs are well-known.  

\section{Spectral Analysis}\label{s:spec}
  
We modeled the spectra of sources with more than $\sim$30 counts between
0.35 and 8.0 keV (147 sources), using the ACIS\_EXTRACT automated
spectral fitting script \citep{Broos02} and XSPEC v. 11.3.  We binned
the data using at least 10 counts per bin, and used $\chi^2$
statistics.  We tried several spectral models, including 
power-law, blackbody, bremsstrahlung, and one-temperature mekal
continuum models \citep{Mewe91} modified by the PHABS photoelectric absorption
model.  In all cases we constrained the hydrogen column to
$\geq1.3\times10^{20}$ cm$^{-2}$, the known cluster value
\citep{Gratton03}.  The model expected to best physically describe cataclysmic
variables and chromospherically active binaries is a thermal plasma
model (comprising a bremsstrahlung continuum plus emission lines),
which we characterize by a variable-abundances mekal model (VMEKAL in
XSPEC). We fix the abundances of this model to appropriate values for
47 Tuc; [Fe/H]=-0.70 (20\% solar), [O/Fe]=0.50 (60\% solar), and
[Si/Fe]=0.30 (40\% solar) \citep{Carney96,
Salaris98}.  We assume C, N and O have similar abundances; Ne through
Ca have similar abundances; and that Fe and Ni have the same abundance.
  We present the results
of our VMEKAL spectral fits to the 2002 data in 
Table~\ref{tab:spec}, and to the 2000 data in
Table~\ref{tab:spec2000}.   Detailed analyses of 
these spectra (including time-resolved spectra for variable sources) and
timing properties of all sources will be the subject 
of additional papers.  In the following we look at some general
properties of the VMEKAL spectral fits to the 2002 data, and briefly
discuss some individual sources.

\subsection{Spectral properties}\label{s:specall}

\begin{figure}
\includegraphics[scale=.80]{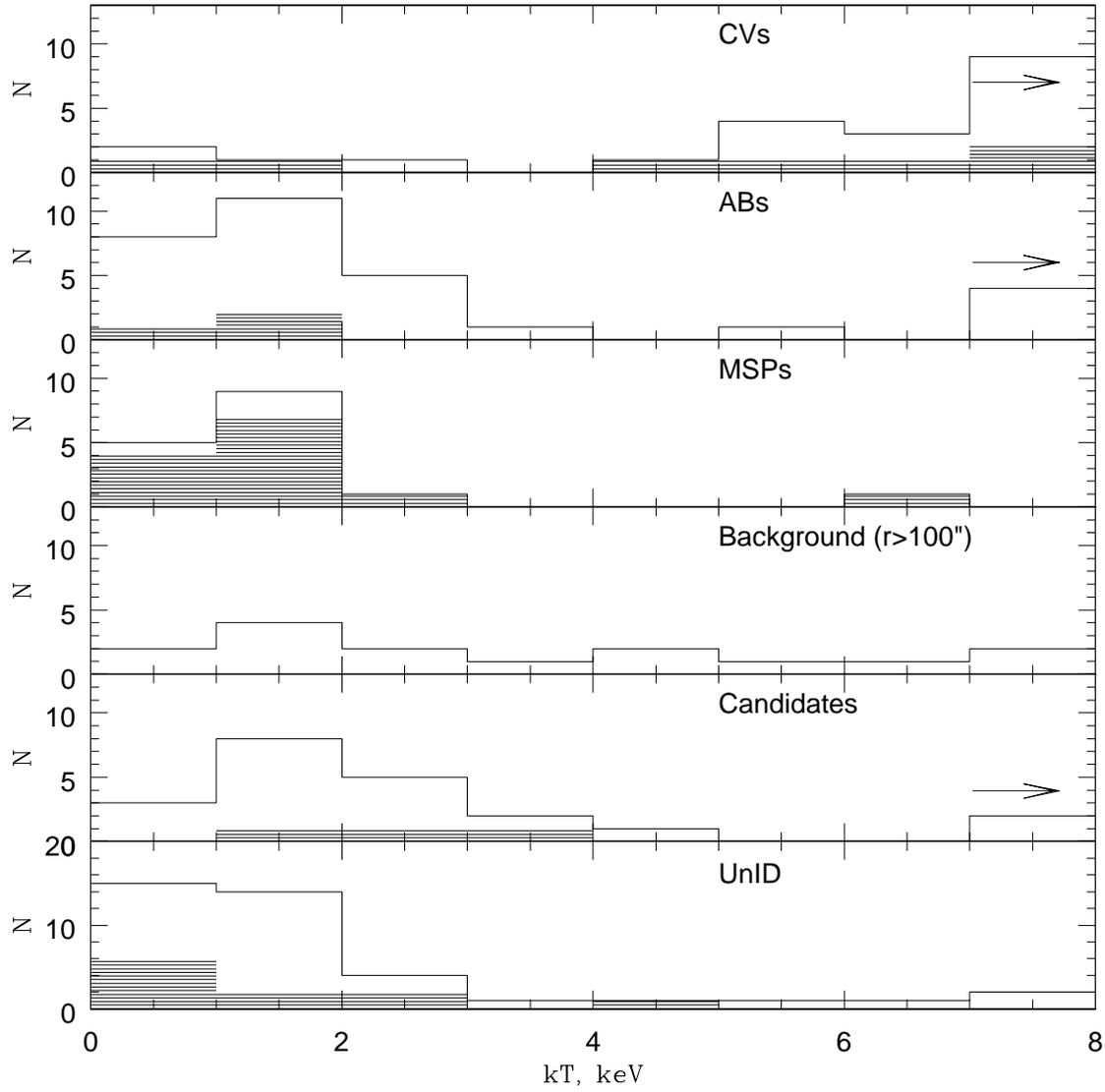}
\caption[kthist.eps]{Histograms of fitted temperatures from thermal
plasma model fits to five source classes.  The group labeled
``Candidates'' includes possible counterparts indicated as 
CV?s, AB?s and MSP?s in \citet{Edmonds03b} and Table~\ref{tab:pos}. All five
qLMXBs and qLMXB candidates give poor fits, and are not shown here.
Bad fits producing null hypothesis probabilities $<$5\% are  
indicated by the shaded histograms.  Any best-fit temperatures above 7
keV  (above which \Chandra has little effective area) 
are included in the last bin. 
} \label{fig:kthist}
\end{figure}

147 sources had sufficient counts for
simple spectral fitting.  Most of the spectra were reasonably well-fit
(null hypothesis probability, or nhp, $>$5\%) by the power-law
(133/147 sources) and bremsstrahlung (129/147 sources) models, while 
the single-temperature VMEKAL model resulted
in slightly fewer (105/147) good fits.  However, the VMEKAL model
returns values of $N_H$ near the known cluster value for most ABs,
unlike the power-law model. 
We anticipate most ABs should not suffer enhanced extinction, and
choose the VMEKAL model as the most physical description for many of
our sources.  We show histograms of the derived temperatures for 
members of various source classes in Figure~\ref{fig:kthist}, along with the
numbers of sources showing poor spectral fits.   Multiple temperature
plasmas may be required to understand some source spectra, while
others are not expected to be well-described by thermal plasmas at all 
(specifically MSPs and qLMXBs).  
We note that most (13/16) of the known MSPs, and all five qLMXBs or
qLMXB candidates, are poorly fit by thermal plasma
models.  One-third (7/21) of the fitted CVs are poorly fit; this is
partly due to their generally very high statistics (exposing
calibration uncertainties) and partly due to complex models required
to describe some of the brighter sources (see below).  Other groups
were better 
fit: only 3/22 ABs, 1/7 candidate MSPs, 2/7 candidate CVs, none of the
15 sources 
beyond 100\arcsec$\!$, and 11/40 of the remaining unidentified sources
were badly fit.  

\begin{figure}
\includegraphics[scale=.80]{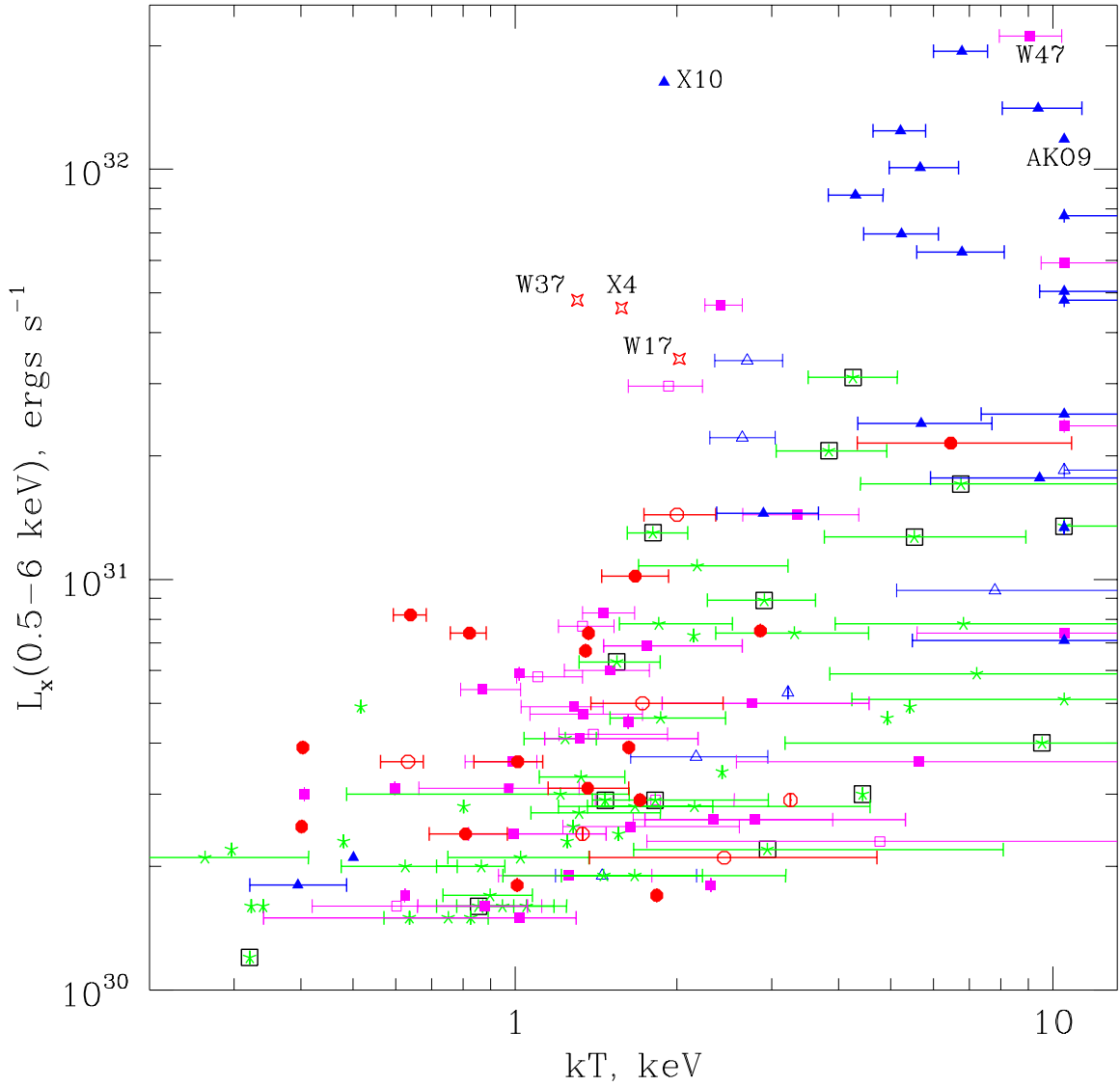}
\caption[sm_kT_bw.eps]{kT from thermal plasma model fit, plotted against
fitted 0.5-6 keV $L_X$.  Symbols as in Figure~\ref{fig:lines}.  Errors (90\%
confidence) on kT are not plotted for sources that are very badly fit
by our thermal plasma model ($\chi^2_{\nu}\geq2.0$), including the
three relatively bright sources X7, X5 and X9 (left off this plot).
Some unusual 
objects are marked. (Please see the electronic edition of the Journal
for a color version of this figure.)
} \label{fig:kTLx}
\end{figure}

We also plot the best-fit temperature against the 0.5-6 keV X-ray
luminosity in Figure~\ref{fig:kTLx}.   We see   
a correlation between temperature and X-ray luminosity
generally followed by both ABs and CVs.  For CVs we expect that both
temperature and luminosity may increase with the white dwarf mass
\citep{Wu95,Cropper98}.  No
correlation of either temperature or luminosity with accretion rate
has been found in previous studies \citep[see][]{vanTeeseling96}.  
  We note that there are
several well-fit CVs with temperatures which clearly lie below 10 keV.
This may imply that their white dwarf masses lie below 0.4 \Msun, or
that they have relatively high rates of cyclotron cooling
\citep{Wu95}.  
The qLMXB candidates appear to have 
unusually low temperatures and high X-ray luminosities compared to other X-ray
sources, and are poorly fit by plasma models (see also below).  

\begin{figure}
\includegraphics[scale=.80]{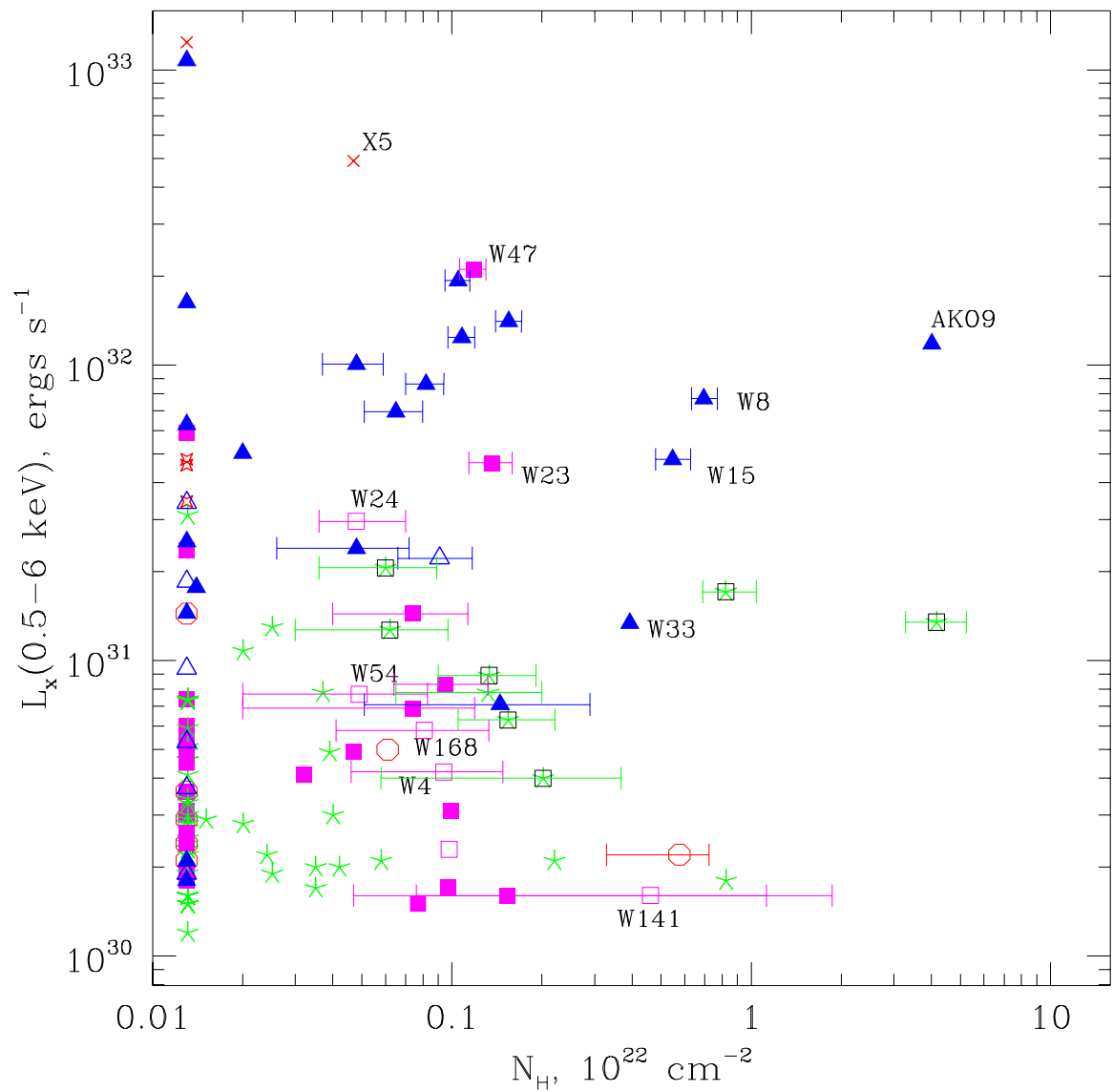}
\caption[sm_Nh_bw.eps]{$N_H$ from thermal plasma model fit, plotted against
fitted 0.5-6 keV $L_X$.  Symbols as in Figure~\ref{fig:lines}.  Errors (90\% 
confidence) on $N_H$ are plotted only for those sources which are
well-fit by this model and which have absorption columns inconsistent
with the cluster value, 
$N_H=1.3\times10^{20}$ cm$^{-2}$ \citep{Gratton03}. Three objects
which are not well-fit by the single thermal plasma model (and thus do 
not have plotted errors), but which
clearly show extra absorption, are identified (X5, AKO9, and
W33).  Four other sources which show unusually high absorption columns 
for their class are also identified, as are the five AB candidates from
\citet{Edmonds03b} with intrinsic absorption (see \S~\ref{s:specall}.
 (Please see the electronic edition of the Journal 
for a color version of this figure.)
} \label{fig:Nh}
\end{figure}

Most of the X-ray sources are consistent with the cluster $N_H$ value
in the VMEKAL fits, but several are clearly not.  We plot the derived
$N_H$ values from our thermal plasma fits in Figure~\ref{fig:Nh}, plotting
errors only where the derived $N_H$ is not consistent with the cluster 
value, and where the fit produces $\chi^2_{\nu}<2.0$.  
Several categories of X-ray sources tend to show enhanced $N_H$; CVs,
likely background sources, and candidate ABs.  The likely background
sources should contain extragalactic AGN, which 
are often highly absorbed, so their $N_H$ values are not surprising.

Twelve of 22 known CVs require intrinsic $N_H$ columns.   The known
eclipsing CVs AKO9, W8, and W15 \citep{Edmonds03b} clearly show very
high $N_H$, as expected for an edge-on inclination; W33 appears to
have a similarly high $N_H$ value, suggesting a similarly high
inclination.  The enhanced $N_H$ values for most of the other bright
CVs, near $10^{21}$ cm$^{-2}$, 
indicate that we are seeing absorption from gas within these systems.
Nonmagnetic systems at low inclination appear not to require enhanced
$N_H$ beyond that expected from interstellar material
\citep{Verbunt97,Eracleous91,vanTeeseling96}.  On the other hand, the
moderately magnetic systems known as intermediate polars or DQ Hers often
show increased $N_H$ \citep{Norton89} and tend to have higher X-ray
luminosities \citep[though this may be affected by selection
effects,][]{Verbunt97}.  It has already been suggested that X-ray
bright CVs in globular clusters may be preferentially magnetic systems
\citep{Grindlay95,Edmonds99,Edmonds03b}, and the enhanced 
$N_H$ values tends to support this conclusion. 

We are surprised by the intrinsic absorption seen in 5 of the
6 candidate ABs identified by \citet{Edmonds03b}, including W24, W54,
W168, W4 and W141.
These systems were identified  
as possible RS CVn due to their proximity to bright stars, subgiants
or giants, which 
were saturated in the GO-8267 \HST\ dataset.  The discovery that
their X-ray spectra are also special indicates that the bright stars
are indeed related to 
these X-ray sources. (If the proximity of these X-ray sources to
bright stars were due to coincidence, then these X-ray sources should
not have a unifying X-ray characteristic.)  
 We speculate that the enhanced $N_H$ may be due to 
dense winds from the giant or subgiant stars.  We have not, however,
confirmed these stars to be RS CVn; other explanations, such as
symbiotic stars (white dwarfs accreting from giant star winds), are still
possible.  These stars are likely to be 
amenable to ground-based optical photometry and spectroscopy, unlike
the majority of X-ray counterparts in 47 Tuc. 

The enhanced $N_H$ from six ABs is also not understood. The brightest, 
W47, is a known semidetached W UMa binary, indicating that Roche lobe
overflow is occurring and may be responsible for the occulting
gas.  Enhanced $N_H$ has been observed for several
chromospherically active stars \citep{Favata99,Maggio00} at the level
of $10^{19}$--$10^{20}$ cm$^{-2}$, and has been attributed to coronal
mass ejections. However, some of our ABs show $N_H\sim10^{21}$
cm$^{-2}$, 10 times higher than observed in field systems.

\subsection{Individual sources}

We briefly discuss a few sources which are poorly fit by a thermal
plasma model, and have unusual values for their best-fitting
parameters.  The two bright qLMXBs X5 
and X7 are very badly fit, partly due to the effects of
pileup on their very soft spectra, which we do not model here
\citep{Heinke03a}. (We note that the modeling of the hard spectra of
the other relatively bright sources is less affected by pileup.)
Hydrogen-atmosphere modeling using the XSPEC pileup formalism provides
excellent fits to these spectra; 
this fitting is discussed in
Rybicki et al. (2005, in preparation).  Most of the rest of the sources can
 be fit by two-component spectra; we discuss the most interesting ones 
below.

\subsubsection{CVs}

\begin{figure}
\includegraphics[scale=.80]{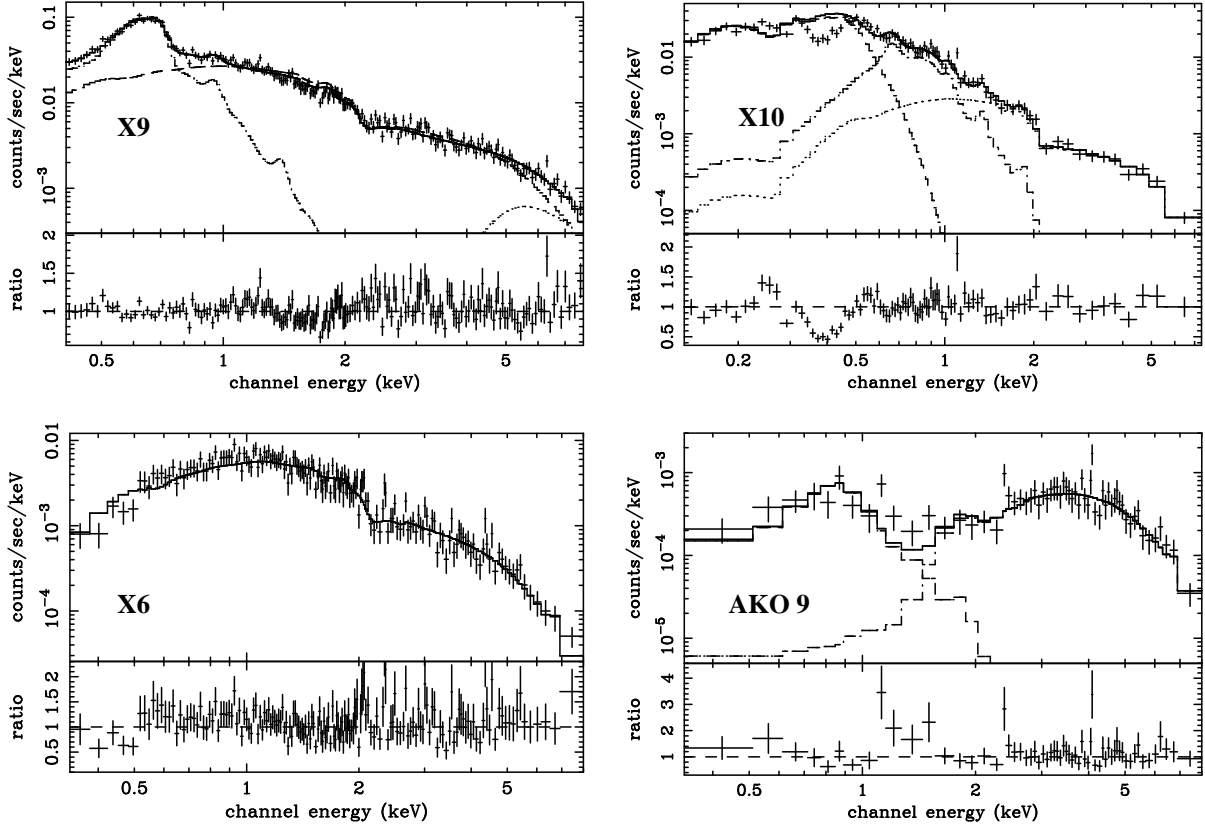}
\caption[CVspec.eps]{Energy spectra of four CVs in 47 Tuc, showing
    their diversity in observational appearance.  {\it Upper
    left:} X9 (W42), a
    bright CV that may be an intermediate polar, modeled with three
    mekal components at 0.25, $>17$, and $>6$ keV.  The third
    component is absorbed
    behind a column of $9\times10^{23}$ H cm$^{-2}$.  {\it Upper
    right:} X10 (W27),
    a probable polar CV, modeled with a kT=53 eV blackbody and two
    mekal plasmas at 0.39 and $>14$ keV.  The feature at 0.4 keV is 
    a known calibration residual. {\it Lower left:} X6 (W56), a
    bright CV in 47 Tuc well-fit with a simple absorbed 
    thermal plasma model, with enhanced $N_H$ over the cluster value.
 {\it Lower  right:} AKO 9 (W36), an eclipsing CV with a
    subgiant secondary. We model 
    the spectrum  with two mekal components at 0.6 and $>5$ keV in
    the 2002 observations.  The second
    component is absorbed behind a column of $5\times10^{22}$ H
    cm$^{-2}$. In 2000, a single plasma component of temperature 0.3
    keV is seen; if the second component is present, it
    is obscured behind 
    $>50\times10^{22}$ H cm$^{-2}$ in the 2000 observations. 
} \label{fig:CVspec}
\end{figure}

The CV X9 (W42), the brightest non-qLMXB in the cluster, shows a
peculiarly strong emission line at 0.65 keV, tentatively identified
with O VIII, on a hard continuum.  To 
model this spectrum requires two mekal components of similar emission
measures, one at 0.25 and one 
at $>$17 keV, with abundances set to those of the cluster.  However, this
does not result in a good fit, primarily because the spectrum contains
so many counts (20,000) that calibration residuals are clearly
apparent. We note, in accord with \citet{Sanders03}, significant
residuals between 1.3 and 2.3 keV, and an overestimate of the flux
around 0.4 keV.  We also see a significant underprediction of the flux
above 6 keV, which cannot be accounted for by pileup, since it is also
present in the spectrum generated from subarray observations. Some of
this excess may be due to calibration errors.  Most of
the excess can be accounted for by adding a third plasma component, hidden
behind a column of 10--100$\times10^{22}$ H cm$^{-2}$.  This would be
consistent with partial covering models often invoked for the
intermediate polar subset of CVs \citep{Norton89}, and the possible
detection of a 218 s period by GHE01a, which they suggested indicated a 
DQ Her classification.  We show this model in Figure~\ref{fig:CVspec}.
In any case, X9's luminosity and
apparent spectral hardness are similar to the Galactic center systems
discussed by \citet{Muno03,Muno04}.

The bright CV X10 (W27) 
exhibits a rather soft spectrum, poorly fit by single component
models.  It is the brightest source in the cluster in the poorly
calibrated 0.1-0.3 keV band, having 1240 counts in that band (compared
to 732 for the bright, soft qLMXB X7), and we believe that correct
understanding of its spectrum cannot be accomplished without including
these soft X-rays.  A spectral
fit using two mekal components (at 0.39$\pm0.03$ and $>14$ keV), plus a
blackbody with $kT=53\pm4$ eV, provides a good fit above 0.5 keV, and
a reasonable fit to most of the data below 0.5 keV
(Fig.~\ref{fig:CVspec}).   
An apparently uncalibrated feature at 0.4 
keV is present in the data from other bright 47 Tuc sources (notably
X7), and in ACIS-S spectra of the Crab Nebula (OBS\_ID 2000).
 X10's spectrum is consistent with those seen from other polars, having
 a soft blackbody-like feature from the white dwarf surface in
 addition to a multitemperature accretion column \citep{Ramsay04}.
 Detailed time-resolved analysis of this source, the first polar CV to be 
identified in a globular cluster, will be presented in Heinke et
al. (2005, in prep).  We note that both CVs X9 and X10 require plasma
components with temperatures above 10 keV, showing that they too may 
follow the empirical relation between X-ray luminosity and $kT$
described above (\S~\ref{s:specall}).  

The CVs W33 and AKO9 (W36) are not well-fit
by mekal or power-law 
spectra.  Both also have hard colors (see \S\ref{s:disc} below), like the
eclipsing absorbed CVs W15 and W8 (GHE01a).  AKO 9,
however, appeared quite soft in the 2000 dataset.  Spectral fits to
AKO 9 require at least two plasma components ($kT=0.6$ and 10 keV), with
 the higher temperature component requiring additional 
absorption by a column of at least $3\times10^{22}$ H cm$^{-2}$.  This
is consistent with AKO 9's known eclipsing behavior \citep{Edmonds96,
  Knigge03}, which implies that we see the accretion impact zone on
the white dwarf \citep{Patterson85} through the edge of the accretion disk. 
The 2000 data shows only one, unabsorbed component, best fit at 0.3
keV; if the second, high-temperature component is present, it must be
even more heavily absorbed (see Fig.~\ref{fig:CVspec}).  W33 also requires
absorption by two different columns; a mekal model with temperature
1.8 keV, 95\% covered with $6\times10^{22}$ H cm$^{-2}$ (the remainder 
covered with $0.3\times10^{22}$ H cm$^{-2}$) gives a good fit.  The
origin of the low-temperature components is unclear; perhaps they are
scattered in an extended region, as observed in other CVs
\citep{Mukai03, Pratt04}.  For comparison with the relatively unusual
spectra of X9, X10 and AKO9, we also show the relatively simple
spectrum of the bright CV X6, fit by a thermal plasma with
$kT=6.8\pm0.8$ keV and $N_H=10.5\pm1.0\times10^{20}$ cm$^{-2}$.  
Additional details on the spectra of all the 47 Tuc CVs will be presented
in future papers. 

\subsubsection{Additional qLMXBs?}

The three sources W37, W17 and X4 (W125) have unusual locations in
our X-ray CMD  (Fig.~\ref{fig:XCMD}), and show poor 
fits (nhp $<$ 5\%) to thermal plasma models (Table~\ref{tab:spec}).  Fitting 
them with two-component thermal plasmas gives an acceptable fit for
X4, but not for the other two.  W17 and X4 are also poorly fit by blackbody, 
bremsstrahlung, and powerlaw models.   These three sources
 are also the three brightest X-ray sources ($L_X\sim4\times10^{31}$
 ergs s$^{-1}$) without known optical 
IDs in 47 Tuc. We note that W17 is suggested as a possible qLMXB by
\citet{Edmonds03a}.  Quiescent LMXBs have been identified at X-ray
luminosities as low as $10^{31}$ ergs s$^{-1}$, and with relatively
hard spectra \citep{Campana02, Tomsick04}.  
Fits of hydrogen-atmosphere models plus power-laws to W17 and X4
 give implied
surface radii ($12^{+10}_{-4}$ and $7^{+4}_{-1}$ km respectively)
consistent with 10-13 km, as expected for neutron stars 
\citep{Lattimer01}, and power-law photon indices of 1.9, similar to those
observed for qLMXBs in the field \citep{Rutledge02b}.  W37's
time-averaged spectrum is also well-fit by a 
hydrogen-atmosphere plus power-law model, but the implied radius is
only $1.7^{+0.6}_{-0.3}$ km.  However, W37 shows some of the most
dramatic temporal and spectral variability in 47 Tuc, and
time-resolved spectra show consistency with a 10 km radius for the
thermal component.  In a companion paper \citep{Heinke04b}, we show
that W37's variability is due to varying $N_H$ and 
eclipses, and argue that W17, X4 and  W37 are qLMXBs.

\section{Discussion}\label{s:disc}

\subsection{MSPs}\label{s:msp}

For the first time, we have clearly detected in X-rays all MSPs in 47 Tuc with
radio timing positions that are not located within 1\arcsec\ of another 
MSP.  This allows us to measure the luminosity function for MSPs in 47
Tuc, describing them as a lognormal distribution with mean
log$L_X=30.6$ and standard deviation log$L_X=0.28$ (0.5-2.5 keV).  

\begin{figure}
\includegraphics[scale=.80]{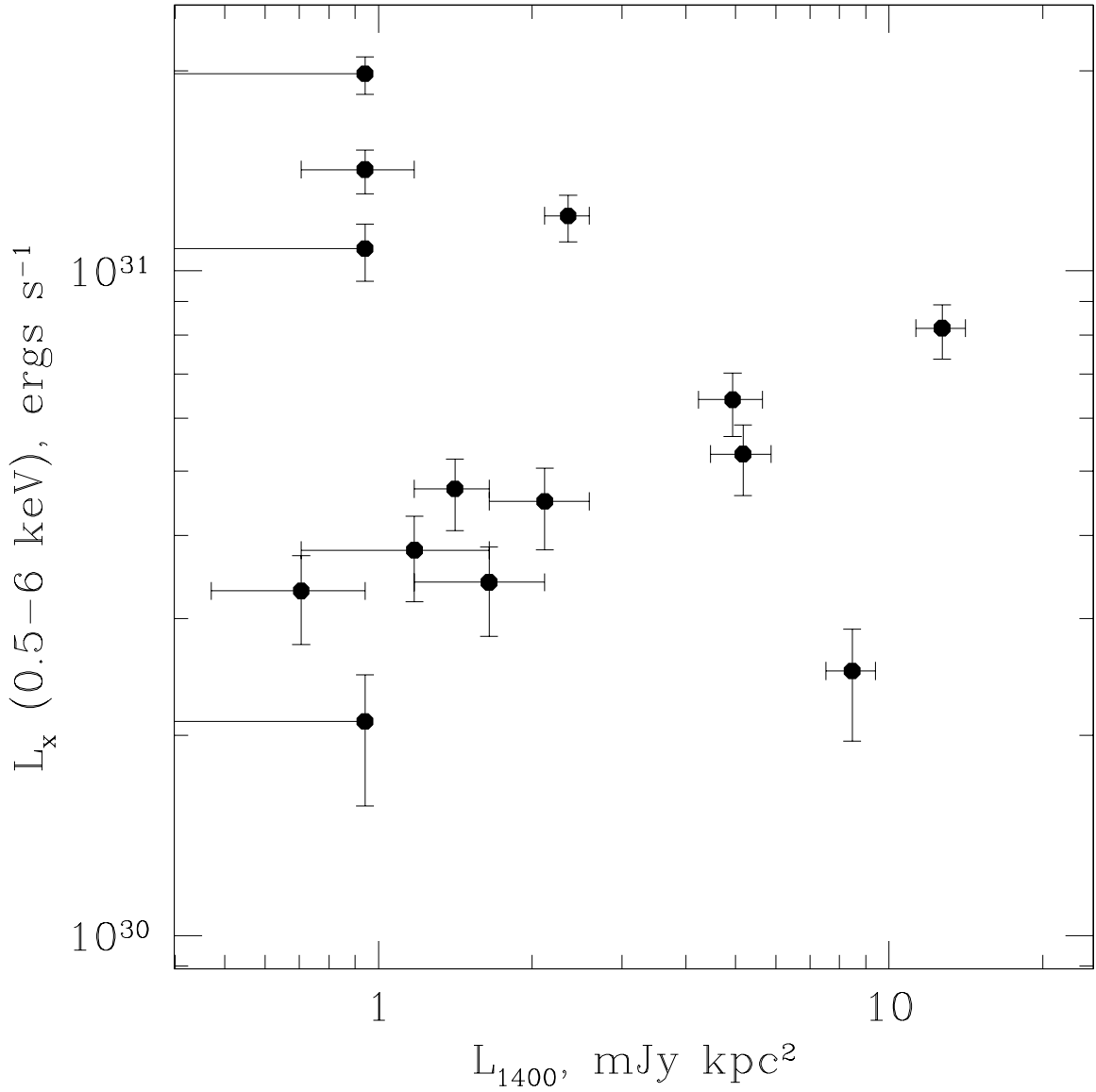}
\caption[sm_MSP_l.eps]{X-ray (0.5-6 keV) and radio (1400 MHz, pseudo-)
  luminosities for individually detected MSPs in 47 Tuc.  Radio
  pseudoluminosities from \citet{Camilo00}, with estimates for MSPs W,
  R, and T (0.04 mJy kpc$^2$) taken as upper limits. No correlation is seen
  between X-ray and radio luminosities.
} \label{fig:MSP_rx}
\end{figure}

We compare the 0.5-2.5 keV and 0.5-6 keV X-ray luminosities (from
Table~\ref{tab:pos}) of the individually  
detected MSPs to their radio pseudoluminosities \citep{Camilo00}, to
look for any 
correlation.  For MSPs which do not have listed pseudoluminosities in
\citet{Camilo00} we use their upper limit estimate of 0.04 mJy kpc$^2$. 
We show the radio pseudoluminosities and X-ray 0.5-6 keV luminosities
in Figure~\ref{fig:MSP_rx}.  Using a Spearman rank-order test
\citep{Press92}, we 
find statistically insignificant negative correlations (rank
correlations $r_s$=-0.05 and -0.01 for 0.5-2.5 and 0.5-6 keV
luminosities respectively, with probabilities of
chance occurrence of a correlation of this strength being 86\% and 98\%). 
This analysis includes rough estimates of the pseudoluminosities of three MSPs
(W, R and T) which do not have measured flux densities listed in
\citet{Camilo00}.  Leaving out these three, we find 
 statistically insignificant positive correlations (rank correlations
 $r_s$=0.07 and 0.14, chance probabilities 82\% and 69\%).

We conclude that we see no correlation of the radio
pseudoluminosities from 47 Tuc MSPs with their X-ray luminosities, in agreement
 with \citet{Grindlay02}.  This is not 
unexpected, since the X-ray emission from 47 Tuc MSPs is largely
thermal radiation from the surface \citep{Grindlay02} while the radio
emission originates in magnetospheric processes above the neutron star.  This lack of any
correlation between X-ray and radio signals indicates that additional
MSPs in 47 Tuc will probably have X-ray luminosities similar to those
MSPs which have already been identified.  Some
MSPs in 47 Tuc may have qualities which make them particularly
difficult to detect 
in the radio, such as extremely short (less than 2 millisecond)
periods, or radio eclipses covering large parts of the orbit.  Very
short-period MSPs might be expected to have relatively large spindown
luminosities, and probably relatively high X-ray luminosities.  Those
MSPs in 47 Tuc which eclipse also tend to have relatively high X-ray
luminosities (W, O and J are among the four X-ray brightest MSPs in 47
Tuc).  It thus seems likely that MSPs with such qualities would be
more, rather than less, X-ray luminous.

Among the 12 candidate MSPs listed by \citet{Edmonds03b}, W99 is
identified as a cosmic ray in the 2000 data, not a real source.  
Four exhibit variability within the 2002 observations
(W6, W96, W31, W91), and the last three of these show  
colors rather different from the majority of the known MSPs.  W97, W95, 
and W115 appear much fainter in 2002
than the known MSPs ($L_X < 1/2$ that of MSP 47 Tuc-T, the faintest X-ray   
MSP known).  This leaves 
W5, W28, W142, W34, and possibly W6 as excellent MSP candidates within 
the GO-8267 \HST\ field and GHE01a source list.  
We identify another group of possible MSP candidates by their failure
to be well-fit by VMEKAL models, together with their location in the
color-color diagram.  These additional MSP candidates are W10 (also a
possible CV candidate), W65, W90, W40, W83, W6, W84, W253, W249, W303,
W200, and W279.  These 12 objects are likely to contain some MSPs, but
may also contain some other objects. 

\subsection{ABs}

It has generally been believed (GHE01a) that ABs only appear as X-ray
sources in globular clusters briefly, during flares, and then
disappear.  This survey indicates that ABs can maintain relatively
high X-ray luminosities over a period of weeks.  One of the brighter
ABs (W41, $L_X>2\times10^{31}$ ergs s$^{-1}$) was not detected as
variable within our observations on any timescale.  Although many ABs do
show flaring behavior, their lightcurves are not easily
distinguishable from those of CVs (see Figure~\ref{fig:lightcurves}).

Thirty-one ABs have been newly identified as X-ray sources in 47
Tuc, and additional ABs will surely be identified using new and
archival \HST\ imaging.  We note that of the eight classes of variables 
in 47 Tuc discussed by AGB01, we have detected 11 of 15 W UMa
binaries, five of six red straggler variables, eight of ten eclipsing
binaries, and 24 of 69 BY Draconis stars as X-ray sources.  X-ray
observations are apparently highly effective at identifying
short-period binaries in globular clusters.

\subsection{Background sources}

In \S\ref{s:radial}, we showed that the distribution of sources with
greater than 
20 counts is consistent with a flat distribution beyond 100\arcsec\ out 
to our 2\farcm79 limiting survey radius.  Thus, these sources are
likely to be background, in that they are not a dynamically relaxed
component of 47 Tuc.  However, in \S\ref{s:lf} we showed that they are
more numerous than the expectations for extragalactic background
sources.  Several explanations are possible, of which the most
interesting is the possibility that some of these sources are associated with
the halo of 47 Tuc.   If they are halo sources, they are probably CVs
and ABs formed by primordial binaries \citep{Davies97,
Gendre03a}. Such a two-component distribution of interacting binaries
has already been inferred for blue stragglers in M3 and 47 Tuc
\citep{Ferraro97, Ferraro04}.   We 
note that only one X-ray source (W211, a faint AB) has been identifed
beyond 100\arcsec\ 
in 47 Tuc.  A future study will examine the sources beyond the half-mass
radius of 47 Tuc, to determine whether a substantial halo population
exists. 

Our sensitivity to faint X-ray sources drops off as we approach the
crowded core of 47 Tuc.  Thus, extrapolating the numbers of detected
sources beyond 100\arcsec\ (55) to the area inside that limit will
overestimate the 
number of background sources we have detected.  A detailed analysis of 
our sensitivity limit through artificial-source simulations, required
to understand the effects of crowding, is beyond
the scope of this work.  We estimate the
total number of X-ray-detected background objects within the half-mass
radius of 47 Tuc to be $\sim$70.  

\subsection{Unidentified sources: Total number of MSPs}\label{s:constraint}

We have identified several differences between the MSPs in 
47 Tuc and the other source classes.  In this section, we attempt to
estimate the makeup of a subset of the unidentified sources by
comparing their properties with those of the 
identified sources, with the goal of constraining the number of MSPs
in 47 Tuc.  We consider 120 sources with between 30 and 350
counts and sufficient spectral bins for spectral fitting, which include all
individually identified MSPs (and no qLMXBs).  
We specify five classes: identified MSPs (16), ABs (27),
CVs (7), likely background sources (those beyond 100\arcsec; 14), and
the remaining unidentified sources (including CV, MSP and AB
candidates; 56).  To constrain the total numbers of a
source population, we require that the identified members were
identified by a process which does not depend on X-ray
characteristics, and that unidentified members have similar X-ray
characteristics (colors, luminosity, variability) to identified
members.  Given the  
measured X-ray luminosities of MSPs, the fact that all MSPs with a known
position were detected in our data, and the lack of correlation between MSP 
X-ray luminosities and radio pseudoluminosities, we expect nearly all
additional unidentified MSPs to fall among our studied subsample, and
to have similar properties to the detected MSPs.  These criteria are not
satisfied by CVs and ABs, so we cannot effectively constrain their
populations by this method; see \S~\ref{s:abcv}.

Our method is based on the assumption that the unidentified sources
 are composed of a combination of sources similar to the identified
 sources,
\begin{equation}
u=k a + l b + m c + n d
\end{equation}
with $u$ being the number of unidentified sources, $a$, $b$, $c$, and $d$ the
 numbers of MSPs, ABs, CVs, and background sources, and $k$, $l$, $m$,
 and $n$ ratios of unidentified to identified sources of each nature.
 We further assume that for each subset $i$ of sources identified by some
 property (such as variability), we can write  
\begin{equation}
u_i \approx k a_i + l b_i + m c_i + n d_i  \label{eq:subset}
\end{equation}
with $u_i$ being the number of unidentified sources sharing a
 property, $a_i$ being the number of known MSPs sharing this property,
 and $k a_i$ being the number of MSPs among the unknown sources
 sharing this property.  Thus we assume that the ratio of the number
 of known MSPs sharing a property to the number of unidentified MSPs
 sharing that property is the same (within errors determined by
 binomial statistics) as the ratio of the total number of known MSPs
 to the total number of unidentified MSPs.

We perform a least-squares fitting of the numbers of objects of
each class to the total number of unidentified objects, within a number of
 groups.  We define these groups by one of the following
properties: the  
sources are variable, or nonvariable; the VMEKAL spectral fit is
adequate, or it is poor (nhp $<$5\%); or the X-ray colors are within one of 
three ranges (HR2$<$0.2 \& HR3$<$-0.5; HR2$>$0.2 \& HR3$<$-0.5;
HR3$>$-0.5).  Each of these groups is subdivided into two subgroups,
including sources with 30--100 counts, and those with 100--350 counts,
making a total of 14 groups.  We define our merit function  
\begin{equation}
\chi^2=\sum_{i=1} \frac{(-u_i+k a_i+l b_i +m c_i+n
d_i)^2}{\sigma_{u_i}^2+k^2 \sigma_{a_i}^2+ l^2 \sigma_{b_i}^2+ m^2
\sigma_{c_i}^2+n^2 \sigma_{d_i}^2}
\end{equation}
where $\sigma_{a_i}$ is the uncertainty on the random variable $a_i$,
computed according to the binomial $1\sigma$ limits of equations 21
and 26 of \citet{Gehrels86}.  We take the relevant rates to be $a_i$
and $a-a_i$ for computation of these binomial errors. 
We vary $n$ within the range $0.56\pm0.11$, 
based on the background number density from \S~\ref{s:radial} and the
geometric areas within and outside 100\arcsec, to 
appropriately scale the numbers of background sources we should
detect. We vary $k$, $l$, $m$ and $n$ to find the lowest value of
$\chi^2$, and define our (enclosed 68\% confidence) errors as
$\Delta\chi^2$=1.  

Forcing the coefficients $k$, $l$, and $m$ to be identical for the
30--100 count bin and the 100--350 count bin, we derive
$k=0.20^{+0.6}_{-0.2}$, $l=1.75^{+0.35}_{-0.35}$, $m=0^{+0.8}$, or
$3^{+10}_{-3}$ additional detected MSPs, $47^{+9}_{-9}$ additional ABs,
and $0^{+6}$ additional CVs within this luminosity range.  
The limit on CVs is not a believable constraint, because we anticipate
that some faint CVs may resemble ABs in their  
X-ray properties (see \S\ref{s:spec}).  Some MSPs may be confused with 
other sources as bright as themselves. This affects the known MSPs
equally (two pairs are 
confused), and thus we multiply our MSP predictions by 1.125 to account 
for undetected sources.  We arrive at $(16+3)\times1.125=22^{+7}_{-4}$
total MSPs in 
47 Tuc, with a 95\% confidence single-sided upper limit of 40 MSPs.

This fit 
effectively incorporates some information about the luminosity
functions, which are not well-determined for CVs and ABs, into the
fitting. We relax the requirement that $k$, $l$ and  
$m$ be equal both for sources with more than 100 counts and those with
less than 100 counts, by performing the fits separately for sources
above and below 100 counts and combining the results.  We find
$30^{+14}_{-12}$ MSPs, 95\% confidence upper limit of 54.  Including
all 132 sources between 30 and 350 counts, and excluding information
about the quality of spectral fits, we find $18^{+15}_{-0}$ MSPs,
95\% confidence upper limit 45, or $27^{+16}_{-9}$ MSPs and upper
limit 57 if $k$, $l$ and $m$ are not held equal for both bins (thus
ignoring both spectral fit and luminosity information).
Considering these various estimates, we suggest 25 as the most likely
number of MSPs in 47 Tuc, and 60 as a conservative upper limit. A
total of 22 MSPs are known through radio studies of 47 Tuc (four
without known positions), providing an independent lower limit on the
number of MSPs in 47 Tuc that is close to our predicted value.

 MSPs that are continually enshrouded in ionized gas from their 
companion and thus invisible to radio searches \citep[similar to 47
 Tuc-V and W,][]{Camilo00,Freire04}, 
or in highly accelerated orbits, or with submillisecond spin periods,
 will be included in this estimate if their X-ray properties are
 similar to those of the known MSPs.   
However, MSPs completely enshrouded in ionized gas may have hard X-ray
 spectra, caused by  
 a shock from the pulsar wind interacting with the gas from the
 companion \citep[as indicated for 47 Tuc-W,][]{Bogdanov05b}.  This
 would alter the X-ray properties of these enshrouded
 MSPs, so they would not be counted in our census.  However, the X-ray
 luminosity from the wind shock must be much larger than the thermal MSP X-ray
 emission to dominate the X-ray spectrum, so such MSPs should be
 relatively bright ($L_X$(0.5-6 keV)$\simge 10^{31}$ ergs s$^{-1}$, like 47
 Tuc-W).  
 There are only 6 unidentified sources within 100$''$ of 47 Tuc with
 $L_X$(0.5-6 keV)$>10^{31}$ ergs s$^{-1}$, which limits the numbers of such
 enshrouded MSPs possible in 47 Tuc.  

Because the X-ray beaming fraction is close to unity \citep{Bogdanov04},
our result constrains the radio beaming fraction to $\gsim37$\% \citep[in
accord with the predictions of][]{Lyne88}.  We
note that our constraint on the total number of MSPs in 47 Tuc is in
 agreement with recent studies 
 using optical identifications of X-ray sources \citep{Edmonds03b},
and limits on the integrated radio flux from the cluster
\citep{McConnell04}, which argue that the total number of MSPs in 47 
Tuc is probably $\lsim30$.  

If we assume that MSPs are subject to the same
scaling with close encounter frequency as LMXBs and qLMXBs
\citep{Verbunt87, Johnston92, Pooley03, Heinke03d}, then our result predicts 
of order 700 MSPs in the Galactic globular cluster system.  This
comes from the fact that 47 Tuc contains 3.6\% of the total close
encounter rate of the Galactic globular cluster system \citep{Heinke03d}.
Predictions of substantially larger numbers of MSPs \citep[e.g., 10,000
in][]{Kulkarni90} have been used to argue that the MSP birthrate is
too high to be explained solely by their formation in LMXB systems.  Our
estimate helps alleviate the discrepancy between LMXB and MSP
birthrates.  We note, however, that another important factor in
reconciling these birthrates may be the prediction that many MSPs
were formed in the distant past through the evolution of
intermediate-mass X-ray binaries \citep{Podsiadlowski02}. 
 
\citet{Edmonds02b} showed that 47 Tuc-W has probably exchanged the
companion which originally spun it up to millisecond periods for
another, main-sequence companion.  \citet{Grindlay02} argued that this 
was also a likely scenario for the eclipsing MSP J1740-5340 in NGC 6397. 
\citet{Freire04} suggested that all 
the six observed eclipsing, low-mass ($M_c \sim 0.1-0.3$ \Msun) binary
pulsars, 
which are only observed in globular clusters, have undergone this
interaction.  (We follow Freire in using low-mass binary pulsars to
mean those with companion masses $>0.1$ \Msun, as opposed to very
low-mass binary pulsars with lower companion masses.)  If this rate of
exchanges is typical of neutron stars in  
general, \citet{Freire04} showed that this implied that about 7.5\% of the
neutron stars in MSP-forming clusters may have been recycled into
MSPs.  This calculation is particularly applicable to 47 Tuc, which
apparently has 
two eclipsing low-mass binary pulsars \citep[47 Tuc-W and probably V]{Freire04}
among 22 known 
radio MSPs.  (It is also supported by recent theoretical work, such as
 \citet{Ivanova04}, which produce recycling rates of
5-13\% in globular cluster models.)  
Dividing our estimate of 25 MSPs in 47 Tuc by 0.075, we
estimate that of order 330 neutron stars exist in 47 Tuc. This number
carries large systematic uncertainties that we conservatively estimate
as a factor of four.  However, we believe
that this is the first estimate driven purely by observations of the
total number of neutron stars in a globular cluster.  This
estimate significantly alleviates the long-standing problem of neutron star
retention in globular clusters \citep{Pfahl02a}, especially when
combined with our current understanding of neutron star kick
distributions \citep{Arzoumanian02, Pfahl02b}.

\subsection{Unidentified sources: CVs vs. ABs}\label{s:abcv}

 Since the physical origin of the X-ray emission is similar for ABs
and CVs (i.e., thermal emission from optically thin gas), it is
difficult to clearly separate them.  The X-ray 
properties of identified CVs also may differ from those of
unidentified CVs, since the known CVs are X-ray selected.

An alternative approach to separating the CVs and ABs is to
extrapolate from the identified objects in the GO-8267 \HST\ field of 
view to the remaining sources. The GO-8267 dataset is the most 
sensitive for detection of both ABs and CVs, due to its extensive
time series information, although several CVs have 
been identified in other \HST\ datasets.  
153 X-ray sources (in the 2002 data) lie in the GO-8267 
field of view, of which 57 are known ABs (26 identified in
\cite{Edmonds03a} and 31 identified here from AGB01 variables), 15
are CVs, 3 are qLMXBs 
(including the qLMXB candidate W17), and 12 are MSPs. Of order 10 may 
be background sources.  In the rest of our field there are 147
sources, of which 60 are likely background sources.   This allows us
to directly extrapolate that 
at least $15/153\times87+15=24$ CVs and $57/153\times87+57=89$ ABs
should be found in 47 Tuc.  Since there 
still remain 66 unidentified sources in the GO-8267 field, these are 
lower limits.  Since the ABs are less centrally concentrated than the
other sources (\S~\ref{s:radial}), and the GO-8267 field encloses the
cluster core, the number of ABs must be even greater.   

Upper limits may be set by assigning the remaining unidentified
non-background X-ray
sources (once the expected MSP contribution has been subtracted) to
either ABs or CVs.  In this way, we restrict the X-ray detected AB
population to $\lsim178$, and the X-ray detected CV population to
$\lsim113$.  This does not account for confusion, nor for objects
not detected in the X-ray; unlike for MSPs, we have reason to believe
that both ABs and CVs may exist at fainter luminosities than our
survey reaches \citep{Verbunt97,Dempsey97}.  Our  
non-detection of the candidate CVs detected in the far-UV by
\citet{Knigge02} also suggests that X-ray faint CVs may lie below our
detection limit. 
Further \HST\ counterpart searches on both new and archival data 
will surely identify additional ABs and CVs among the detected X-ray
sources.  Only 17 of 77 real X-ray sources from the 2000 \Chandra
dataset in the GO-8267 \HST\ field of view
remain unidentified (22\%), when the 5 AB candidates showing enhanced
$N_H$ (see
\S~\ref{s:specall}), 6 MSPs with radio positions but no optical
counterparts, and 3 qLMXBs are considered.  We anticipate, therefore,
that at least half the new unidentified sources in the GO-8267 field may
be identified using new and archival \HST\ data. 

\section{Conclusions}

We have detected 300 X-ray sources within the half-mass radius of 47
Tuc in a deep 281 ks \Chandra ACIS-S observation taken in late 2002,
reaching down to $L_X < 8\times10^{29}$ ergs s$^{-1}$.  All 18 MSPs
with known positions in 47 Tuc are detected, although two closely
spaced pairs of MSPs are unresolved.  All 
but three of 146 sources identified in the 72 ks \Chandra ACIS-I
observations in 2000 are detected in the new observations.  
We confidently identify thirty-one X-ray sources with ABs from the optical
variable lists  
of AGB01, in addition to the 22 CVs and 29 ABs optically identified by
\citet{Edmonds03a}.  
Seventy-eight X-ray sources show evidence of variability, at the 99\%
confidence level or higher, on timescales from hours to weeks within
the 2002 dataset.  Nine additional sources show variability between
the 2000 and 2002 observations.  

 Based on the radial distributions of the X-ray sources, roughly 70
sources are attributed to a background population that is not
concentrated around the cluster center.  The luminosity function for the
observed background population implies an additional source population 
besides the expected extragalactic contribution. The radial profile of the
sources associated with the cluster core allows an estimate of the
mass ratio as $1.63\pm0.11$ $M_*$, where $M_*$ is the average mass of
the stars used to determine the core radius in \citet{Howell00}, and
thus an inferred average system mass of $M=1.43\pm0.17$ \Msun. This is
consistent with our expectations for
binary and/or heavy X-ray sources.  The AB population is
found to be significantly less massive, with $M=0.99\pm0.13$ \Msun\
implied.  The MSP radial distribution implies an average system mass of
$M=1.47\pm0.19$ \Msun.  Subtracting the known average mass of the
binary counterparts gives an average neutron star mass of
$1.39\pm0.19$ \Msun, indicating that these MSPs have accreted very
little mass during their recycling.

The X-ray colors and spectra of the X-ray sources are generally consistent
with those 
expected from a thermal plasma model, in some cases requiring
increased $N_H$ over the known cluster value.  The MSPs are generally not
consistent with a thermal plasma model.  Their colors suggest that
their spectra 
lie between the predictions of blackbody and power-law models.  The
spectra of both the CVs and ABs tend to become harder as their X-ray
luminosity increases.  The bright CVs tend to show increased $N_H$ over 
the cluster value.  Some of these CVs are known to be eclipsing (and
thus high inclination), but the prevalence of high $N_H$ among bright
CVs may suggest a magnetic DQ Her nature for the bright systems.  The X-ray
sources identified by \citet{Edmonds03b} as possible ABs (on the basis
of proximity to bright subgiant or giant stars which are saturated
in the \HST\ data) are unusual in that
5/6 of them require high $N_H$ values, indicating that the bright stars are
indeed associated with the X-ray sources.

The two previously identified qLMXBs in 47 Tuc display the same
spectral and variability  
characteristics as in the 2000 dataset \citep{Heinke03a}. We identify three
additional likely qLMXBs (W37, W17, and X4=W125).  These three objects 
stand out due to their soft colors and unusual spectra, and require a
spectral component that resembles a neutron star surface
\citep{Heinke04b}.  The bright CV known as X9 exhibits a strong 0.65
keV emission line, likely O VIII. We are not aware of
another spectrum of a CV showing such a strong flux from such a low
temperature (0.25 keV) component. The CV X10 shows strong flux below
0.5 keV, and sinusoidal 
modulations with a 16.8 ks period, identifying it as a magnetic AM Her
system (the first confirmed in a globular cluster). 
Using the numbers of detected CVs and ABs in the deeply studied
GO-8267 \HST\ field, we constrain the numbers of CVs and ABs among
our detected 47 Tuc sources to 24-113 CVs, and 89-178 ABs.  Additional 
CVs and ABs probably lie below our detection limit.  

No correlation is found between the radio and X-ray luminosities of
MSPs in 47 Tuc, motivating us to use the properties of detected MSPs
to constrain the population of undetected MSPs, which we assume are
similar.    
The difference in colors between MSPs and objects dominated by
optically thin plasma radiation, along with variability and spectral
fitting information, can be used to constrain the total
number of MSPs in 47 Tuc to $22^{+7}_{-4}$ (95\% confidence upper
limit of $\sim$60), regardless of their
radio beaming fraction.  This allows us to estimate the total number
of MSPs in the galactic globular cluster system (assuming formation in 
the same manner as LMXBs) as $\sim$700.  If the eclipsing low-mass
 ($\ga 0.1$ \Msun\ companion) binary pulsar systems are good tracers
for the likelihood of multiple 
recycling episodes for neutron stars, then we can estimate the total
number of neutron stars in 47 Tuc as $\sim$300.

Future papers will probe the spectra, variability, radial
distributions, and optical properties of each of the
classes of sources in 47 Tuc in greater detail.  Our new simultaneous
\HST\ ACS data, along with archival \HST\ data, will also enable many
additional optical identifications. 

\acknowledgements

We thank the Penn State team, particularly Pat Broos, for the
development and support of the ACIS\_EXTRACT software.  
We thank D. Lloyd for the use of his neutron
star atmosphere models, and M. Albrow for results of his \HST\
radial completeness calculations. We also thank R. Wijnands and
R. di~Stefano for useful discussions, R. Narayan for useful comments on the
manuscript, and the referee for a careful and helpful report.  This
work has made substantial use of NASA's Astrophysics Data 
System, and was supported in part by Chandra grant GO2-3059A.
C.~H. also acknowledges the support of the Lindheimer fund at
Northwestern University.

Facilities: {CXO(ACIS)}.

\bibliography{src_ref_list}
\bibliographystyle{apj}

% figures
\clearpage

\clearpage

\clearpage

%tables

\clearpage

\begin{deluxetable}{lccccr}
\tablewidth{5.5truein}
\tablecaption{\textbf{Summary of \Chandra Observations}}
\tablehead{
\colhead{Seq,OBS\_ID} & \colhead{Start Time} & \colhead{Exposure} &
\colhead{Aimpoint} & \colhead{Frametime} & \colhead{CCDs}
}
\startdata
300003,078 & 2000 Mar 16 07:18:30  &  3875  & ACIS-I & 0.94 & 1/4 \\
300028,953 & 2000 Mar 16 08:39:44  & 31421  & ACIS-I & 3.24 & 6 \\
300029,954 & 2000 Mar 16 18:03:03  &   845  & ACIS-I & 0.54 & 1/8 \\
300030,955 & 2000 Mar 16 18:33:03  & 31354  & ACIS-I & 3.24 & 6 \\
300031,956 & 2000 Mar 17 03:56:23  &  4656  & ACIS-I & 0.94 & 1/4 \\
400215,2735 & 2002 Sep 29 16:59:00 & 65237 & ACIS-S & 3.14 & 5 \\
400215,3384 & 2002 Sep 30 11:38:22 &  5307 & ACIS-S & 0.84 & 1/4 \\
400216,2736 & 2002 Sep 30 13:25:32 & 65243 & ACIS-S & 3.14 & 5 \\
400216,3385 & 2002 Oct 01 08:13:32 &  5307 & ACIS-S & 0.84 & 1/4 \\
400217,2737 & 2002 Oct 02 18:51:10 & 65243 & ACIS-S & 3.14 & 5 \\
400217,3386 & 2002 Oct 03 13:38:21 &  5545 & ACIS-S & 0.84 & 1/4 \\
400218,2738 & 2002 Oct 11 01:42:59 & 68771 & ACIS-S & 3.14 & 5 \\
400218,3387 & 2002 Oct 11 21:23:12 &  5735 & ACIS-S & 0.84 & 1/4 \\
\enddata
\tablecomments{Subarrays are indicated by fractional numbers of CCDs. 
} \label{tab:obs}
\end{deluxetable}

\clearpage

\LongTables

\begin{landscape}
\begin{deluxetable}{lcccccccccccr}
\tablecolumns{13}
%\rotate
\tablewidth{9.8truein}
\tabletypesize{\footnotesize}
\tablecaption{\textbf{47 Tuc Basic X-ray Source Properties, 2002 Dataset}}
\tablehead{
\multicolumn{2}{c}{Names} & \multicolumn{2}{c}{Position} & & 
\multicolumn{3}{c}{Counts} & \multicolumn{2}{c}{$L_X$} & & &  \\
\colhead{W} & \colhead{CXO GLB J} & \colhead{$\alpha$ (J2000)} &
\colhead{$\delta$ (J2000)}
& \colhead{Dist.} & \colhead{0.5-4.5} & \colhead{0.5-1.5}  &
\colhead{1.5-6}  & \colhead{0.5-6.0}  & \colhead{0.5-2.5}  &
\colhead{ID} & \colhead{Other} & \colhead{Notes} \\
& & (h m s err) & ($\circ$ $'$ \arcsec\ err) & (\arcsec) & keV & keV &
keV & \multicolumn{2}{c}{($10^{30}$ ergs s$^{-1}$)} & {Type} & {Names} & \\
(1) & (2) & (3) & (4) & (5) & (6) & (7) & (8) & (9) & (10) & (11) &
(12) & (13) 
} 
\startdata
46 & 002403.4-720451 & 00 24 03.494 0.001 & -72 04 51.94 0.01 &     8 & 31407.2$^{+179.1}_{-177.2}$ & 26426.7$^{+164.0}_{-162.6}$ & 4987.6$^{+72.1}_{-70.6}$ & 1760.1$^{+10.0}_{-9.9}$ & 1227.1$^{+7.0}_{-7.0}$ & qLX & X7 & - \\ 
42 & 002404.2-720458 & 00 24 04.249 0.001 & -72 04 58.02 0.01 &     7 & 16661.4$^{+131.0}_{-129.1}$ & 11569.7$^{+109.0}_{-107.6}$ & 5954.6$^{+78.9}_{-77.2}$ & 1107.5$^{+8.5}_{-8.4}$ & 588.2$^{+4.9}_{-4.9}$ & CV & X9 & DHY \\ 
58 & 002400.9-720453 & 00 24 00.947 0.001 & -72 04 53.20 0.01 &    20 & 11130.1$^{+107.4}_{-105.5}$ & 8723.8$^{+94.8}_{-93.4}$ & 2422.4$^{+50.7}_{-49.2}$ & 581.7$^{+5.6}_{-5.5}$ & 429.2$^{+4.2}_{-4.1}$ & qLX & X5 & DHY \\ 
27 & 002406.3-720443 & 00 24 06.374 0.001 & -72 04 43.00 0.01 &    10 & 3947.1$^{+64.7}_{-62.8}$ & 3241.3$^{+58.4}_{-56.9}$ & 815.1$^{+30.3}_{-28.5}$ & 259.3$^{+4.2}_{-4.1}$ & 147.8$^{+2.5}_{-2.4}$ & CV & X10 & DH \\ 
47 & 002403.4-720505 & 00 24 03.450 0.002 & -72 05 05.28 0.01 &    15 & 2629.8$^{+53.3}_{-51.3}$ & 1375.2$^{+38.5}_{-37.1}$ & 1387.7$^{+38.9}_{-37.2}$ & 147.6$^{+2.9}_{-2.8}$ & 85.5$^{+1.9}_{-1.9}$ & AB & - & DHY \\ 
56 & 002402.1-720542 & 00 24 02.119 0.002 & -72 05 42.03 0.01 &    51 & 2590.4$^{+52.9}_{-50.9}$ & 1487.5$^{+40.0}_{-38.6}$ & 1222.9$^{+36.7}_{-35.0}$ & 143.5$^{+2.9}_{-2.8}$ & 86.2$^{+1.9}_{-1.9}$ & CV & X6 & DHY \\ 
1 & 002416.9-720427 & 00 24 16.968 0.002 & -72 04 27.21 0.01 &    59 & 1810.6$^{+44.5}_{-42.5}$ & 1080.1$^{+34.3}_{-32.9}$ & 804.7$^{+30.1}_{-28.3}$ & 97.3$^{+2.4}_{-2.2}$ & 59.7$^{+1.6}_{-1.5}$ & CV & - & H?Y \\ 
2 & 002415.8-720436 & 00 24 15.880 0.002 & -72 04 36.38 0.01 &    51 & 1714.0$^{+43.4}_{-41.4}$ & 846.7$^{+30.5}_{-29.1}$ & 956.7$^{+32.6}_{-30.9}$ & 96.1$^{+2.4}_{-2.3}$ & 55.8$^{+1.5}_{-1.5}$ & CV & X13 & DHY \\
\enddata
\tablecomments{ Basic properties of sources in the 2002 47 Tuc X-ray
dataset. See \S~\ref{s:extract} for details.  Notes indicate variability or confusion: H and D indicate
hours- and days-timescale variability at 
  99.9\% confidence, H?  and D? indicate possible short- and long-term
  variability at 99\% confidence, and Y indicates 99.9\% confidence
  variability on timescales of years.
A c indicates a possibly confused
source, while an m indicates a source added manually. 
}\label{tab:pos}
\end{deluxetable}
\clearpage

\begin{deluxetable}{lccccccccccccr}
\tablecolumns{14}
%\rotate
\tablewidth{9.2truein}
\tabletypesize{\footnotesize}
\tablecaption{\textbf{47 Tuc Basic X-ray Source Properties, 2000 Dataset}}
\tablehead{
\multicolumn{2}{c}{Names} & \multicolumn{2}{c}{Position} & & &
\multicolumn{3}{c}{Counts} & \multicolumn{2}{c}{$L_X$} & & &  \\
\colhead{W} & \colhead{CXO GLB J} & \colhead{$\alpha$ (J2000)} &
\colhead{$\delta$ (J2000)}
& \colhead{$f_{PSF}$} & \colhead{$f_{EXP}$} & \colhead{0.5-4.5} & \colhead{0.5-1.5}  &
\colhead{1.5-6}  & \colhead{0.5-6.0}  & \colhead{0.5-2.5}  &
\colhead{ID} & \colhead{Other} & \colhead{Notes} \\
& & (h m s err) & ($\circ$ $'$ \arcsec\ err) & &  & keV & keV &
keV & \multicolumn{2}{c}{($10^{30}$ ergs s$^{-1}$)} & {Type} & {Names} & \\
(1) & (2) & (3) & (4) & (5) & (6) & (7) & (8) & (9) & (10) & (11) &
(12) & (13) 
} 
\startdata
46 & 002403.4-720451 & 00 24 03.489 0.001 & -72 04 51.92 0.01 & 0.96 & 1.00 & 5308.2$^{+74.7}_{-72.8}$ & 4403.7$^{+67.8}_{-66.4}$ & 904.4$^{+31.4}_{-30.1}$ & 1338.0$^{+18.8}_{-18.4}$ &  1185.4$^{+16.8}_{-16.4}$ & qLX & X7 & - \\ 
58 & 002400.9-720453 & 00 24 00.940 0.001 & -72 04 53.18 0.01 & 0.96 & 1.00 & 4424.2$^{+68.4}_{-66.5}$ & 3573.3$^{+61.2}_{-59.8}$ & 854.8$^{+30.6}_{-29.2}$ & 1113.6$^{+17.2}_{-16.7}$ &  958.9$^{+15.0}_{-14.6}$ & qLX & X5 & DHY \\ 
42 & 002404.2-720458 & 00 24 04.242 0.002 & -72 04 57.98 0.01 & 0.94 & 1.00 & 2407.3$^{+51.1}_{-49.0}$ & 1470.3$^{+39.8}_{-38.3}$ & 1165.7$^{+35.9}_{-34.1}$ & 684.4$^{+13.9}_{-13.3}$ &  442.0$^{+10.5}_{-10.1}$ & CV & X9 & H?Y \\ 
56 & 002402.1-720542 & 00 24 02.111 0.004 & -72 05 41.95 0.02 & 0.96 & 1.00 & 725.8$^{+29.0}_{-26.9}$ & 353.4$^{+20.2}_{-18.8}$ & 415.3$^{+22.1}_{-20.4}$ & 195.2$^{+7.6}_{-7.0}$ &  109.9$^{+4.9}_{-4.5}$ & CV & X6 & Y \\ 
27 & 002406.3-720443 & 00 24 06.370 0.003 & -72 04 42.97 0.01 & 0.95 & 0.99 & 727.5$^{+28.9}_{-26.9}$ & 562.4$^{+25.1}_{-23.7}$ & 194.9$^{+15.7}_{-13.9}$ & 193.1$^{+7.6}_{-7.0}$ &  162.9$^{+6.8}_{-6.4}$ & CV & X10 & DH \\ 
30 & 002406.0-720456 & 00 24 05.991 0.004 & -72 04 56.18 0.03 & 0.92 & 1.00 & 320.4$^{+20.0}_{-17.9}$ & 176.1$^{+14.7}_{-13.2}$ & 158.1$^{+14.3}_{-12.5}$ & 89.1$^{+5.5}_{-4.9}$ &  51.8$^{+3.6}_{-3.3}$ & CV & X19 & Y \\ 
45 & 002403.7-720423 & 00 24 03.756 0.005 & -72 04 22.91 0.02 & 0.95 & 0.99 & 258.0$^{+18.1}_{-16.0}$ & 148.4$^{+13.6}_{-12.2}$ & 123.5$^{+12.9}_{-11.1}$ & 69.2$^{+4.8}_{-4.2}$ &  43.4$^{+3.3}_{-2.9}$ & CV & - & - \\ 
1 & 002416.9-720427 & 00 24 16.963 0.006 & -72 04 27.18 0.03 & 0.95 & 0.99 & 249.4$^{+17.9}_{-15.7}$ & 131.7$^{+12.9}_{-11.4}$ & 133.6$^{+13.3}_{-11.5}$ & 67.4$^{+4.7}_{-4.1}$ &  41.7$^{+3.2}_{-2.9}$ & CV & - & Y \\
\enddata
\tablecomments{Properties of the 47 Tuc X-ray sources, from 
the 2000 dataset.  Notes indicate variability or confusion: these are
as in Table~\ref{tab:pos}, except that a t indicates a possibly
transient (or spurious) source not seen in the 2002 data.
}\label{tab:pos2000}
\end{deluxetable}

\clearpage
\end{landscape}

\begin{deluxetable}{lccccccr}
\tablewidth{6.6truein}
\tablecaption{\textbf{New Optical Counterpart Identifications}}
\tablehead{
\colhead{W} & \colhead{CXOGLB J} & \colhead{AGB01 name} & \colhead{$\Delta$ R.A.}
& \colhead{$\Delta$ Decl.} & \colhead{Sep./error} & \colhead{Type} &
\colhead{Per.~(d)} 
} 
\startdata
287 & 002405.4-720457 & PC1-V21 & -0.665 & 0.765 & 2.87 & N.Ecl. S.Det. & 0.39 \\
266 & 002407.9-720502 & PC1-V10 & 0.165 & 0.01 & 1.42 & W UMa S.Det. & 0.43 \\
299 & 002404.1-720512 & PC1-V30 & 0.101 & 0.023 & 1.08 & BY Dra & 0.80 \\
267 & 002407.8-720458 & PC1-V04 & 0.036 & 0.023 & 0.41 & W UMa Con. & 0.33 \\
286 & 002405.6-720452 & PC1-V32 & 0.073 & 0.162 & 2.00 & BY Dra & 1.64 \\
304 & 002402.0-720502 & PC1-V42 & 0.345 & 0.382 & 3.43  &  BY Dra & 0.52 \\
274 & 002407.1-720452 & PC1-V34 & -0.065 & -0.049 & 0.97 &  BY Dra & 1.38 \\
273 & 002407.3-720504 & PC1-V18 & -0.092 & 0.162 & 1.48  & W UMa Con. & 0.30 \\
\enddata
\tablecomments{Separations ($\Delta$ R.A. and $\Delta$ Decl.) are in arcseconds, and defined as 
\Chandra minus (shifted) \HST\ positions. Sep./error is the RA and
Decl. separations divided by the \Chandra WAVDETECT errors plus systematic
\HST\ plate solution errors, added in
quadrature. Type indicates classification by
AGB01 (on the basis of lightcurves) as detached eclipsing binaries, BY Dra
variables, W UMa contact 
or semidetached systems, noneclipsing semidetached systems, or red
stragglers.  Periods from AGB01.
}\label{tab:ids}
\end{deluxetable}

\begin{landscape}
\clearpage

\begin{deluxetable}{lccccccccccr}
\tablecolumns{12}
%\rotate
\tabletypesize{\footnotesize}
\tablewidth{8.3truein}
\tablecaption{\textbf{47 Tuc Supplemental X-ray Source Properties, 2002 Dataset}}
\tablehead{
\multicolumn{2}{c}{Names} & \multicolumn{4}{c}{Counts} & & 
\multicolumn{4}{c}{Hardness Ratios} &  \\
\colhead{W} & \colhead{CXOGLB J} & $f_{PSF}$ & $f_{EXP}$ & \colhead{0.3-8.0} &
\colhead{0.3-0.8} & \colhead{0.8-2.0} & \colhead{2.0-8.0} & \colhead{HR1} &
\colhead{HR2} & \colhead{HR3} & \colhead{HR4}  \\
 & & & &  keV & keV & keV & keV & & & &  \\
(1) & (2) & (3) & (4) & (5) & (6) & (7) & (8) & (9) & (10) & (11) &
(12) 
} 
\startdata
46 & 002403.4-720451 & 0.96 & 1.00 & 33887.0$^{+186.2}_{-184.1}$ & 12716.4$^{+114.1}_{-112.8}$ & 19329.1$^{+140.4}_{-139.0}$ & 1841.5$^{+44.0}_{-42.9}$ & -0.83$^{+0.00}_{-0.00}$ & 0.21$^{+0.01}_{-0.01}$ & -0.75$^{+0.01}_{-0.01}$ & 5.30$^{+0.08}_{-0.08}$  \\ 
42 & 002404.2-720458 & 0.94 & 1.00 & 19486.9$^{+142.1}_{-139.6}$ & 7562.4$^{+88.3}_{-87.0}$ & 7325.6$^{+87.0}_{-85.6}$ & 4598.9$^{+69.4}_{-67.8}$ & -0.23$^{+0.01}_{-0.01}$ & -0.02$^{+0.01}_{-0.01}$ & -0.24$^{+0.01}_{-0.01}$ & 1.94$^{+0.03}_{-0.03}$  \\ 
58 & 002400.9-720453 & 0.96 & 1.00 & 11692.0$^{+110.3}_{-108.1}$ & 3204.8$^{+57.9}_{-56.6}$ & 7652.4$^{+88.8}_{-87.5}$ & 834.8$^{+30.2}_{-28.9}$ & -0.80$^{+0.01}_{-0.01}$ & 0.41$^{+0.01}_{-0.01}$ & -0.59$^{+0.01}_{-0.01}$ & 3.60$^{+0.08}_{-0.08}$  \\ 
27 & 002406.3-720443 & 0.95 & 1.00 & 5463.6$^{+76.3}_{-73.9}$ & 3201.9$^{+58.0}_{-56.6}$ & 1725.7$^{+42.9}_{-41.5}$ & 536.0$^{+24.8}_{-23.1}$ & -0.53$^{+0.02}_{-0.02}$ & -0.30$^{+0.01}_{-0.01}$ & -0.71$^{+0.01}_{-0.01}$ & 3.98$^{+0.16}_{-0.16}$  \\ 
47 & 002403.4-720505 & 0.95 & 1.00 & 2878.3$^{+56.1}_{-53.6}$ & 379.8$^{+20.9}_{-19.5}$ & 1569.2$^{+41.0}_{-39.6}$ & 929.3$^{+32.0}_{-30.5}$ & -0.26$^{+0.02}_{-0.02}$ & 0.61$^{+0.02}_{-0.02}$ & 0.42$^{+0.03}_{-0.03}$ & 0.99$^{+0.04}_{-0.04}$  \\ 
56 & 002402.1-720542 & 0.96 & 1.00 & 2806.8$^{+55.4}_{-53.0}$ & 410.3$^{+21.6}_{-20.2}$ & 1587.8$^{+41.2}_{-39.8}$ & 808.7$^{+30.0}_{-28.4}$ & -0.33$^{+0.02}_{-0.02}$ & 0.59$^{+0.02}_{-0.02}$ & 0.33$^{+0.03}_{-0.03}$ & 1.22$^{+0.05}_{-0.05}$  \\ 
1 & 002416.9-720427 & 0.95 & 1.00 & 1955.0$^{+46.6}_{-44.2}$ & 309.6$^{+19.0}_{-17.6}$ & 1145.1$^{+35.2}_{-33.8}$ & 500.2$^{+23.9}_{-22.3}$ & -0.39$^{+0.02}_{-0.02}$ & 0.57$^{+0.02}_{-0.02}$ & 0.24$^{+0.04}_{-0.03}$ & 1.34$^{+0.06}_{-0.06}$  \\ 
2 & 002415.8-720436 & 0.95 & 1.00 & 1860.6$^{+45.6}_{-43.1}$ & 210.9$^{+15.9}_{-14.5}$ & 1039.3$^{+33.7}_{-32.2}$ & 610.4$^{+26.3}_{-24.7}$ & -0.26$^{+0.02}_{-0.02}$ & 0.66$^{+0.02}_{-0.02}$ & 0.49$^{+0.03}_{-0.03}$ & 0.89$^{+0.04}_{-0.04}$  \\
\enddata
\tablecomments{Additional properties of the 47 Tuc X-ray sources, from 
the 2002 dataset. See \S~\ref{s:extract} for details.  
}\label{tab:hr}
\end{deluxetable}

\clearpage
\end{landscape}

\begin{deluxetable}{lccccr}
\tablewidth{5.3truein}
\tablecaption{\textbf{ROSAT Flux Estimates and Variability}}
\tablehead{
\multicolumn{2}{c}{\Chandra names} & \colhead{ROSAT name} & \colhead{Class}
& \colhead{ROSAT $L_x$} & \colhead{Variable?} 
} 
\startdata
W46 & 002403.4-720451 & X7 & qLX & 1450.4$\pm80.6$ &  - \\ 
W42 & 002404.2-720458 & X9 & CV & 919.7$\pm68.7$ &   T,V,D,H,Y  \\ 
W58 & 002400.9-720453 & X5 & qLX & 1076.8$\pm56.8$ &   T,D,H,Y  \\ 
W27 & 002406.3-720443 & X10 & CV & 236.1$\pm25.1$ &   T,V,D,H  \\ 
W56 & 002402.1-720542 & X6 & CV & 77.9$\pm11.9$ &  D,H,Y \\ 
W1 & 002416.9-720427 & - & CV & $<$31.7 &  T,H?,Y  \\ 
W2 & 002415.8-720436 & X13 & CV & 31.7$\pm7.9$ & D,H,Y  \\ 
W125 & 002353.9-720350 & X4 & qLX? & 64.7$\pm11.9$ &  D?,H \\ 
W25 & 002407.1-720545 & X11 & CV & 51.5$\pm10.6$ &  D \\ 
W30 & 002406.0-720456 & X19 & CV & 153.1$\pm19.8$ &   T,D,Y  \\ 
\enddata
\tablecomments{X-ray sources detected by ROSAT in 47 Tuc, and one 
  X-ray source detected by \Chandra (W1) which would have been detected by
  ROSAT if it had been at its 2002 flux level in the 1990s.  ROSAT
  0.5-2.5 keV X-ray luminosities, in units of $10^{30}$ ergs s$^{-1}$,
  are derived from the source counts of \citet{Verbunt98} 
  assuming a 2 keV Raymond-Smith thermal plasma spectrum with the
  cluster absorption column and a distance of 4.85 kpc.  T indicates
  sources which varied between 
  the 1990s and 2002 (timescale of tens of years), while V indicates
  sources which varied during the 1990s \citep{Verbunt98}.  D, H, and
  Y indicate variability within the 2002 observations, or between the
  2000 and 2002 observations (see Table~\ref{tab:pos}).
}\label{tab:rosvar}
\end{deluxetable}

\clearpage

% Use Spec_tab4.dat, or later version.  
\begin{deluxetable}{lccccccccr}
\tablecolumns{10}
%\rotate
%\tablewidth{9.2truein}
\tabletypesize{\footnotesize}
\tablecaption{\textbf{Spectral Fits to 47 Tuc Sources}}
\tablehead{
\multicolumn{2}{c}{Names} & &
  \multicolumn{5}{c}{VMEKAL Fits} &  \\ 
W & CXOGLB J & Counts 
& $N_H$ & kT & EM & $L_{X,abs}$ & $L_{X,unabs}$ & $\chi^2_{\nu}/\nu$ & Notes \\
 & & 0.3-8 keV  & 
 $10^{20}$ cm$^{-2}$ & keV & $10^{54}$ cm$^{-3}$ &
 \multicolumn{2}{c}{($10^{30}$ ergs s$^{-1}$)} &  &  \\
(1) & (2) & (3) & (4) & (5) & (6) & (7) & (8) & (9) & (10) \\
} 
\startdata
46 & 002403.4-720451 & 33887.0 & 1.3 & 1.1 & 150.56 & 1238.82 & 1284.00 & 17.61/195 & - \\ 
 42 & 002404.2-720458 & 19486.9 & 1.3 & 6.2 & 85.59 & 1076.15 & 1095.00 & 8.89/378 & DHY \\ 
 58 & 002400.9-720453 & 11692.0 & 4.7 & 1.3 & 64.58 & 491.86 & 550.59 & 4.41/169 & DHY \\ 
 27 & 002406.3-720443 &  5463.6 & 1.3 & 1.9 & 18.28 & 163.13 & 167.66 & 13.08/144 & DH \\ 
 47 & 002403.4-720505 &  2878.3 & 11.8$^{+1.2}_{-1.2}$ & 9.0$^{+1.4}_{-1.1}$ & 18.02$^{+0.48}_{-0.47}$ & 211.18$^{+6.71}_{-10.51}$ & 238.50 & 1.09/174 & DHY \\ 
 56 & 002402.1-720542 &  2806.8 & 10.5$^{+1.0}_{-1.0}$ & 6.8$^{+0.8}_{-0.8}$ & 16.88$^{+0.46}_{-0.45}$ & 193.56$^{+8.94}_{-11.48}$ & 218.00 & 1.12/170 & DHY \\ 
 1 & 002416.9-720427 &  1955.0 & 10.8$^{+1.1}_{-1.1}$ & 5.2$^{+0.6}_{-0.6}$ & 11.42$^{+0.39}_{-0.39}$ & 124.17$^{+6.98}_{-6.49}$ & 141.65 & 1.24/133 & H?Y \\ 
 2 & 002415.8-720436 &  1860.6 & 15.5$^{+1.6}_{-1.5}$ & 9.4$^{+1.9}_{-1.3}$ & 12.34$^{+0.41}_{-0.39}$ & 140.93$^{+5.33}_{-9.99}$ & 163.79 & 1.12/132 & DHY \\
\enddata
\tablecomments{Spectral fits to the 47 Tuc X-ray sources from the
  2002 observations.   Spectral fits use an
absorbed thermal plasma VMEKAL model, with 
  abundances set to those of 47 Tuc (see \S~\ref{s:spec}).  When
  $\chi^2_{\nu}>2$, no errors are calculated for spectral parameters;
  otherwise $1\sigma$ errors are quoted.  $L_{X,abs}$ is the inferred
0.5-6 keV X-ray luminosity without accounting for absorption,
  and $L_{X,unabs}$ is the de-absorbed 0.5-6 keV X-ray luminosity
implied by the fitted $N_H$ (and 
  should be treated with skepticism).  
  Notes indicate variability or confusion, which may affect the spectral
  parameters: H and D indicate hours- and days-timescale variability at
  99.9\% confidence, H?  and D? indicate possible hours- and
  days-timescale variability at 99\% confidence, and Y indicates
  years-timescale variability.
A c indicates a possibly confused
source, while an m indicates a source added manually. 
}\label{tab:spec}
\end{deluxetable}

\clearpage

\begin{deluxetable}{lccccccccr}
\tablecolumns{10}
%\rotate
%\tablewidth{9.2truein}
\tabletypesize{\footnotesize}
\tablecaption{\textbf{Spectral Fits to 47 Tuc Sources, 2000 Observations}}
\tablehead{
\multicolumn{2}{c}{Names} & &
  \multicolumn{5}{c}{VMEKAL Fits} &  \\ 
W & CXOGLB J & Counts 
& $N_H$ & kT & EM & $L_{X,abs}$ & $L_{X,unabs}$ & $\chi^2_{\nu}/\nu$ & Notes \\
 & & 0.3-8 keV  & 
 $10^{20}$ cm$^{-2}$ & keV & $10^{54}$ cm$^{-3}$ &
 \multicolumn{2}{c}{($10^{30}$ ergs s$^{-1}$)} &  &  \\
(1) & (2) & (3) & (4) & (5) & (6) & (7) & (8) & (9) & (10) \\
} 
\startdata
46 & 002403.4-720451 & 5308.2$^{+74.7}_{-72.8}$ & 1.3 & 1.0 & 125.35 & 1035.37 & 1074.40 & 4.84/122 &  - \\ 
58 & 002400.9-720453 & 4428.4$^{+68.5}_{-66.5}$ & 1.3 & 1.1 & 108.99 & 895.86 & 928.23 & 3.30/122 &  DHY \\ 
42 & 002404.2-720458 & 2787.7$^{+55.2}_{-52.8}$ & 1.3 & 63.4 & 85.39 & 915.30 & 926.71 & 2.75/174 &  H?Y \\ 
56 & 002402.1-720542 & 791.5$^{+30.5}_{-28.1}$ & 13.7$^{+4.2}_{-4.0}$ & 8.6$^{+3.1}_{-1.9}$ & 23.14$^{+1.51}_{-1.34}$ & 265.70$^{+18.23}_{-30.15}$ & 305.39 & 1.07/65 &  Y \\ 
27 & 002406.3-720443 & 769.1$^{+30.0}_{-27.7}$ & 1.3 & 2.6 & 15.59 & 155.06 & 158.82 & 3.86/60 &  DH \\ 
30 & 002406.0-720456 & 336.1$^{+20.7}_{-18.3}$ & 7.9$^{+4.3}_{-3.8}$ & 5.7$^{+1.9}_{-1.2}$ & 9.66$^{+0.93}_{-0.83}$ & 110.51$^{+13.12}_{-21.16}$ & 121.90 & 0.81/29 &  Y \\ 
45 & 002403.7-720423 & 276.7$^{+19.1}_{-16.6}$ & 1.3$^{+4.2}_{-0.0}$ & 7.2$^{+3.1}_{-2.1}$ & 6.60$^{+0.60}_{-0.41}$ & 84.50$^{+6.47}_{-20.82}$ & 85.92 & 1.28/21 &  - \\ 
1 & 002416.9-720427 & 270.1$^{+18.9}_{-16.4}$ & 5.4$^{+5.2}_{-4.1}$ & 8.1$^{+5.5}_{-2.4}$ & 6.93$^{+0.70}_{-0.62}$ & 85.51$^{+10.98}_{-25.50}$ & 91.12 & 1.14/23 &  Y \\
\enddata
\tablecomments{Spectral fits to 47 Tuc X-ray sources from the
  2000 observations.  Only 32 sources are bright enough for spectral
modeling, and all of these fits are presented here.  Spectral fits use an
absorbed thermal plasma VMEKAL model, with 
  abundances set to those of 47 Tuc (see \S~\ref{s:spec}).  Columns as in
  Table~\ref{tab:spec}.  
}\label{tab:spec2000}
\end{deluxetable}

\end{document}